\documentclass[final]{article}
\usepackage{arxiv}
\usepackage{graphicx,cite}
\usepackage{textcomp}
\usepackage{subcaption}
\usepackage{mathtools}
\usepackage[ruled,vlined]{algorithm2e}
\usepackage{algorithmicx}
\usepackage{booktabs}
\usepackage{amssymb}
\usepackage{pifont}
\usepackage[]{tabto}
\usepackage{verbatim}
\usepackage{ragged2e}
\usepackage{algpseudocode}
\usepackage[]{tabto}
\usepackage{verbatim}
\usepackage{caption}
\usepackage{xcolor}
\usepackage{stfloats}
\graphicspath{{Images_PUF-Authentication-for-IoV/}}
\usepackage{graphicx}
\usepackage{wrapfig}
\usepackage{multirow}
\usepackage{hyperref}
\usepackage{amsmath}
\usepackage{caption}
\usepackage{subcaption}
\makeatletter
\newcommand*{\rom}[1]{\expandafter\@slowromancap\romannumeral #1@}

\begin{document}

	\title{Easy-Sec: PUF-Based Rapid and Robust Authentication Framework for the Internet of Vehicles}

\author{
	Pintu Kumar Sadhu\\
	School of Engineering \& Technology\\
	Central Michigan University\\
	sadhu1pk@cmich.edu\\
	\and
	\textbf{Venkata P. Yanambaka}\\
	School of Engineering \& Technology\\
	Central Michigan University\\
	yanam1v@cmich.edu\\
	\and
	\textbf{Saraju P. Mohanty}\\
	Dept. of Computer Science and Engineering\\
	University of North Texas, USA\\
	saraju.mohanty@unt.edu
	\and
	\textbf{Elias Kougianos}\\
	Dept. of Electrical Engineering\\
	University of North Texas, USA\\
	elias.kougianos@unt.edu
}

\maketitle

\begin{abstract}
With the rapid growth of new technological paradigms such as the Internet of Things (IoT), it opens new doors for many applications in the modern era for the betterment of human life. One of the recent applications of the IoT is the Internet of Vehicles (IoV) which helps to see unprecedented growth of connected vehicles on the roads. The IoV is gaining attention due to enhancing traffic safety and providing low route information. One of the most important and major requirements of the IoV is preserving security and privacy under strict latency. Moreover, vehicles are required to be authenticated frequently and fast considering limited bandwidth, high mobility,  and density of the vehicles. To address the security vulnerabilities and data integrity, an ultralight authentication scheme has been proposed in this article. Physical Unclonable Function (PUF) and XOR function are used to authenticate both server and vehicle in two message flow which makes the proposed scheme ultralight, and less computation is required. The proposed Easy-Sec can authenticate vehicles maintaining low latency and resisting known security threats. Furthermore, the proposed Easy-Sec needs low overhead so that it does not increase the burden of the IoV network. Computational ( around 4 ms) and Communication (32 bytes) overhead shows the feasibility, efficiency, and also security features are depicted using formal analysis, Burrows, Abadi, and Needham (BAN) logic, and informal analysis to show the robustness of the proposed mechanisms against security threats.

\end{abstract}
	
%


\section{Introduction} 
\label{SEC:Introduction}
With the rapid advancement and development of the IoT, both people and devices are being interconnected now-a-days. The IoT has become an inseparable part of our daily life by making integration of physical and digital communities. This trends and development are evidencing \emph{industry 4.0} which was coined in 2011. The forth industrial revolution will bring advancement in the manufacturing process by successive prediction, control, maintenance, and integration of physical build (such as sensors, actuators, complex machinery etc.) and cyber parts (such as network, software etc.). Industry 4.0 is categorized as: IoT, cyberphysical systems, fog computing, cloud computing, Big Data analytics, robotics, augmented and virtual reality, cybersecurity to semantic web technologies, and additive manufacturing \cite{9540626}. Industry is doing huge investment and according to Gartner report, by 2021 devices consisting sensors and actuators will make spend around \$2.5 million \cite{8843960}.  Furthermore, according to the Gartner report \cite{yin2019efficient,Gartner}, the IoT is one of the top ten fastest growing and emerging technology with the largest business potential as there will be 20 times more smart devices by 2023. It also suggested that the revenue of the IoT market will be declined by 3 \% in 2020 which will be rebounded in 2021. In the era of industrial 4.0, for creating beneficial impacts the IoT is transforming and revolutionizing into a wide range of fields, such as the IoV, wearables to robots, smart city, urban planning, power, and so on \cite{liang2020behavioral}. Among these applications, the IoV is gaining attention of academia, researchers, industrialists, and so on. Over the last few years in vehicular network and road, the number of autonomous vehicles (AV) is rising sharply in general, and it is estimated that it may cross two billion within the next 10-20 years \cite{jia2015survey}. The increment trend takes part in the introduction of the intelligent transportation system and related technologies that can enable services, such as smart road sign management, traffic management, efficient insurance policy, toll collection, passenger management, etc. According to a report Association for Safe International Road Travel, nearly 1.25 million people die in road crashes each year which makes an average of 1,287 deaths a day. Another report shows that 1.3 million deaths every year and over 50 million people are injured in car accidents and it is possible to avoid 60 \% - 70 \% accidents \cite{zadobrischi2021vehicular}. Human error is the cause in 90\% of cases. The IoV is a potential solution to enhance the driving experience by improving convenience and road safety \cite{mejri2014survey}. As every devices are smart and also connected, the network has been targeted to become under attack such as cyber and physical attacks. The cyber attack on Ukraine's power infrastructure caused outages of 225,000 consumers in 2015. Also, regular cyber attacks are attempted on U.S. power grids and other systems. So, network of the IoT or the industry 4.0 such as the IoV demands robust and secure system \cite{8843960}. Table \ref{TABLE:Acronyms} shows acronyms and symbols are used in in this article. 

\begin{table}[h!]
	\centering
		\caption{Acronyms and Symbols used in the Current Paper}
	\begin{tabular} {l l}
		Notation & Description  \\ [0.5ex]
		\hline
		Easy-Sec & PUF-Based Rapid and Robust Authentication Framework \\
		IoV &  Internet of Vehicles \\
		VANET & Vehicular ad-hoc Network \\
		V2X & Vehicle to Everything\\
		V2V & Vehicle-to-Vehicle \\
		V2I & Vehicle-to-Infrastructure\\
		V2S & Vehicle-to-Sensors \\
		V2N & Vehicle-to-Network\\
		V2P & Vehicle-to-Pedestrian\\
		AV &  Autonomous Vehicle \\
		RSU &  Roadside Unit \\
		RG &  RSU Gateway \\
		CS &  Cloud Server 	\\
		SDB &  Secure Database 	\\
		PUF & Physical Unclonabale Function \\
		$C$ &  Challenge  	\\
		$R$ & Response \\
				$V_{PID}$ &  Pseudo Identity of AV (PUF) \\
		$PID_{New}$ &  New Pseudo Identity of AV (PUF) \\
		$N_{v}$, $N_{s}$ &  Random Nonce \\
		$R_{C}$ &  Response for Challenge C 	\\
		$R_{C+I}$ &  Response for Challenge (C+I) 	\\
		$R^{K}$ &  K-bit Response for Challenge C 	\\
		$(Y)_R$ & Encryption of message Y using key R \\
		$R (Y)$ & Decryption of message Y using key R \\
		$K, I$ &  Random Integer Number 	\\
		$Z$ &  Function of Random Integer Number 	\\
		$ || $ & Concatenation	\\
		$ \oplus $ & XOR Operation	\\
		$\in$ & Store Operation \\
		$F_{nl}$() &  Any Non-linear Function \\
		\hline
	\end{tabular}

	\label{TABLE:Acronyms}
\end{table}

\subsection{Internet of Vehicles}
\begin{figure}[htbp]
	\centering
	\includegraphics[width=0.75\textwidth]{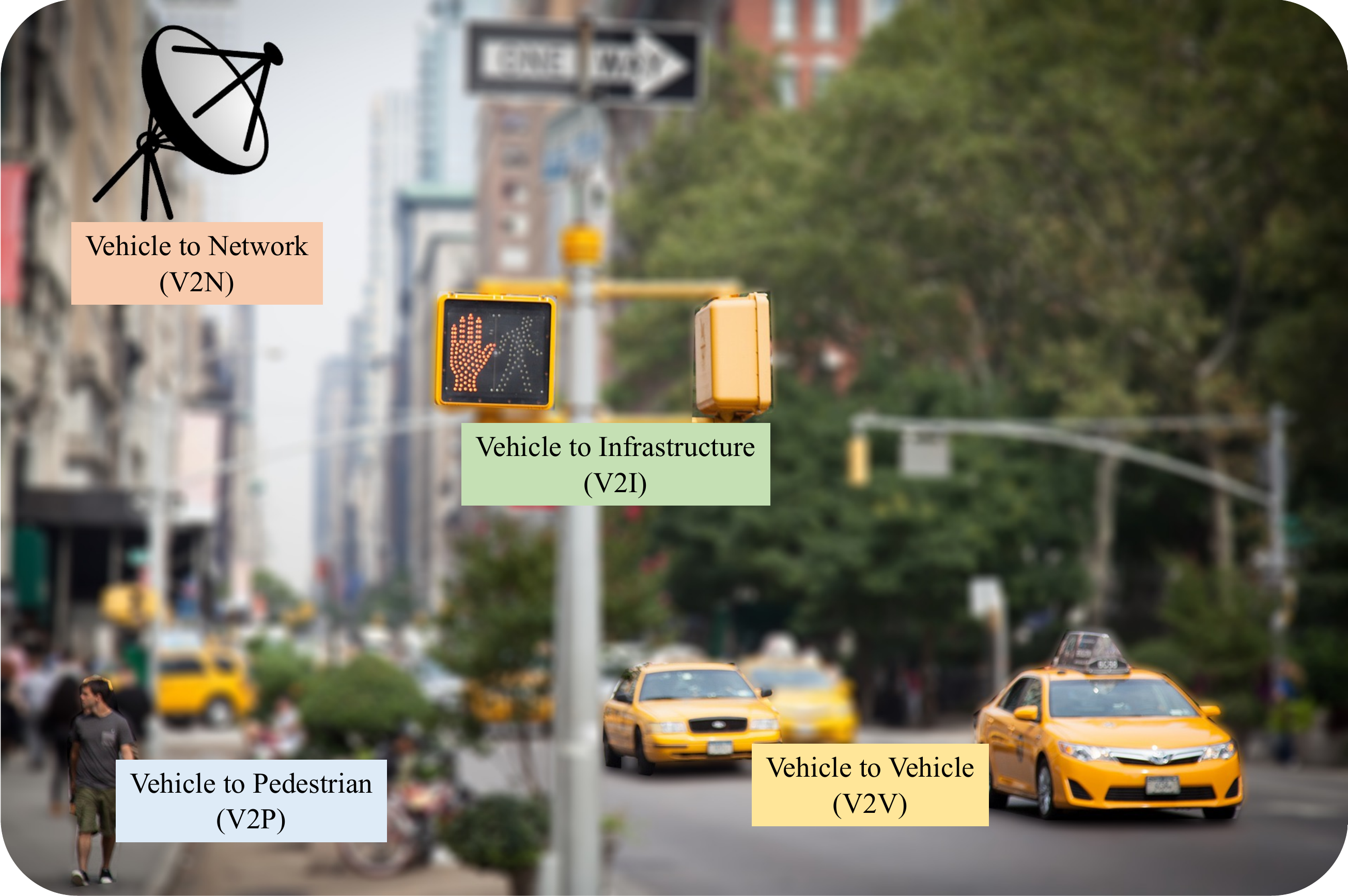}
	\caption{Components in an IoV or V2X Network}
	\label{FIG:V2X}
\end{figure}

The IoV is a massive network, next generation vehicular network coupling the potential of the VANETs and the IoT, which works as a communication hub for people, cars, and various IoT devices that are part of the traffic infrastructure. Not only hardware, software, and services, but also various network technologies,  such as, Bluetooth\textregistered and cellular to Wi-Fi and 5G, as well as different types of communication (V2V, V2X, etc.) are part of the IoV ecosystem \cite{sherazi2019heterogeneous}. Moreover, the incorporation of the IoV and next generation network promotes the V2X \cite{schmidt2016public}. It mainly consists of V2V, V2S, V2I, V2N, and V2P communication \cite{ji2020survey} \cite{haque2020lora} \cite{karami2020smart}. Fig. \ref{FIG:V2X} represents the V2X or the IoV network.

\begin{itemize}
	\item \textit{Vehicle-to-vehicle communication (V2V)}: V2V
	represents inter AV communication on the road which intends to avoid collisions, reduce travel time; with the help of V2V, autonomous driving plays a significant role in increasing road safety for passengers.
	\item \textit{Vehicle-to-sensors communication (V2S)}: V2S or intra-AV is designed as the closed network. In modern cars, the are a number of electronic control units (ECU) \cite{gu2016security} which ensures that a car can communicate with itself (modern AVs are facilitated by various types of sensors for different purposes) and provides an internal performance review. 
	\item \textit{Vehicle-to-infrastructure communication (V2I)}: It is also required to collect data on road infrastructures to make proper analysis for efficient traffic signal control to avoid congestion so that users follow low traffic routes, for avoiding collision, and many more. V2I allows AVs to communicate with roadside units (RSU) to exchange data.
	\item \textit{Vehicle-to-network communication (V2N)}:
	In V2N communication, AVs communicate with cloud or more networks, servers, or data centers for accessing relevant stored information via APIs in the network. Researchers need to focus more on this part of V2X communication.
	\item \textit{Vehicle-to-pedestrian communication (V2P)}:
	In the V2P communication system, AVs communicate with humans on the road such as people and cyclists to detect movement and gather relevant information to avoid accidents. Smartphones and wearable sensor devices, etc. play a great role in setting up V2P communication. 
\end{itemize}

Considering current trend and growth, researchers forecast that approximately 8 million autonomous or semi-autonomous cars will be on the road by 2025. In order to provide clear understanding to industry, government, and people, The Society of Automotive Engineers (SAE) defines 6 levels of driving automation ranging from fully manual (level 0) to fully autonomous (level 5) \cite{SYNOPSYS}. 

\begin{itemize}
	\item \textit{Level 0 - No Automation}: The driver will be responsible for control wheel steers, brakes, accelerators, and follow traffic.
	\item \textit{Level 1 - Driver Assistance}: In this level, car can control single automated system such as steering and braking. Drivers need to monitor so that in case of any failure. s/he can take control. Adaptive cruise control is a good example of this level.
	\item \textit{Level 2 - Partial Automation}: In certain circumstances, the vehicle can steer, accelerate and brake. The driver needs to be in the seat to follow traffic and change lanes. Tesls's autopilot, Cadillac's super cruise etc. falls under level 2.
	\item \textit{Level 3 - Conditional Automation}: In this level, vehicle can manage most features of driving as well as detect environmental conditions. In case of any emergency, it will driver to intervene and take control. An example of this level is 2019 Audi A8.
	\item \textit{Level 4 - High Automation}: Self driving mode is available in level 4. The vehicle can intervene if something is wrong or system failure. However, driver can override the autonomous system. Alphabet's Waymo has a level 4 self-driving taxi service.
	\item \textit{Level 5 - Full Automation}: In the car, user needs to put destination point and car will move without human intervention. The car do not even has steering wheel or acceleration/braking pedals.
\end{itemize}

\subsection{Security Mechanisms}
\begin{figure}[htbp]
	\centering
	\includegraphics[width=.65\textwidth]{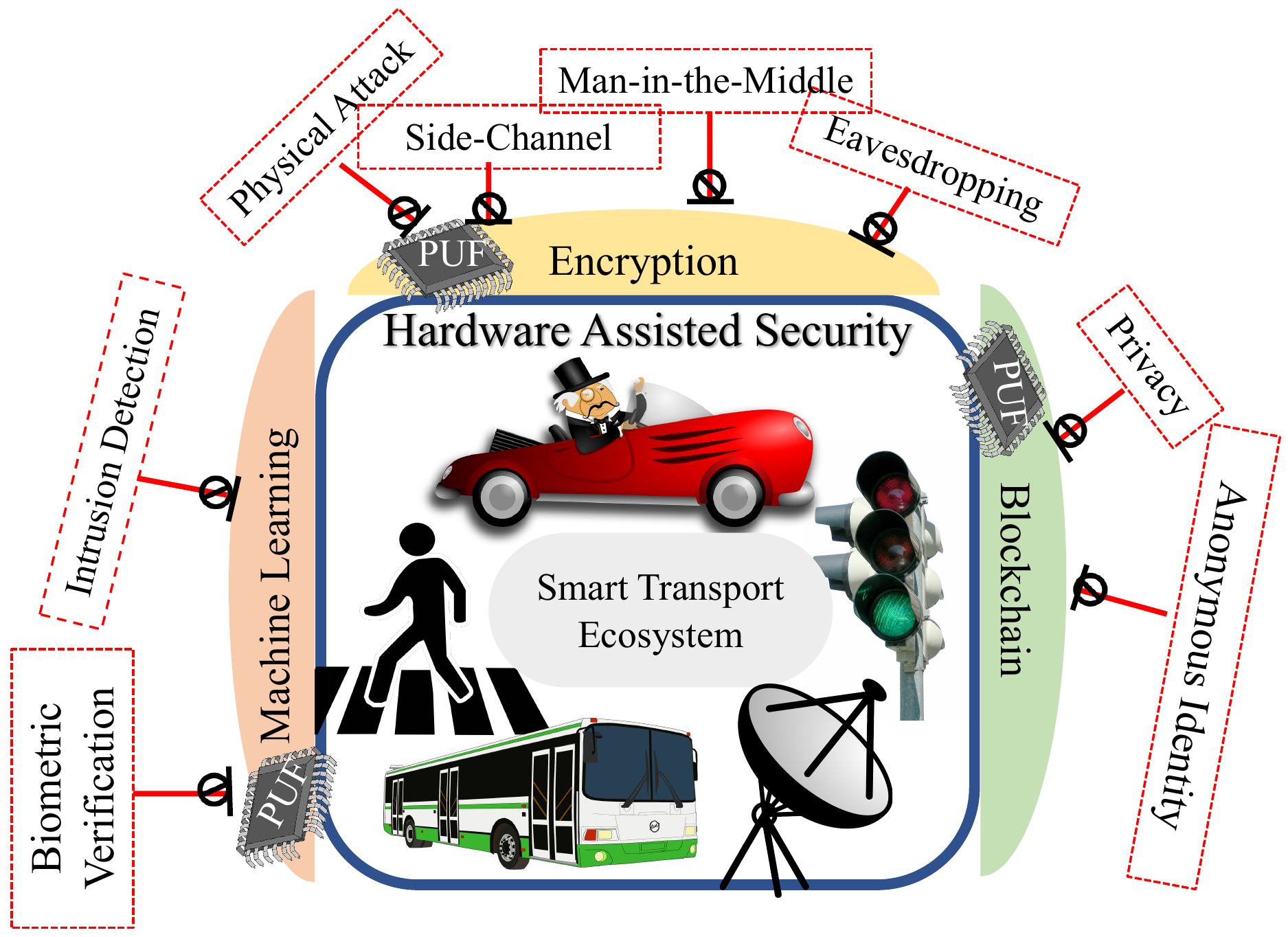}
	\caption{Hardware Assisted Security Mechanisms for Smart Transportation Ecosystem}
	\label{FIG:Security_Mechanisms}
\end{figure}

The ECUs in AVs control one or more electronic control systems with the help of a controller area network (CAN) \cite{gu2016security}. Though it is considered absolutely secured, security issues are introduced due to multinetwork access and fusion. Moreover, with the sheer introduction of vehicles to the internet, the IoV security is going to be a major addressed issue in the near future \cite{alladi2020consumer}. The report ``Cybersecurity Best Practices for Modern Vehicles" of ``National Highway Traffic Safety Administration (NHTSA)" states that mitigation of cybersecurity of transport system is the highest priority of the United States Department of Transportation (DOT) to preserve sensitive information of consumers and ensure safety \cite{thapliyal2019emerging}. To focus on the security area of the IoV network, different technologies and protection schemes, such as in \cite{hassija2020dagiov} \cite{chaudhary2019best} \cite{alladi2019blockchain}, are being explored and experimented. To resist security threats, both software-based and hardware-based mechanisms can be taken. Though it is tough to break the mathematical algorithm of software, but with the help of quantum computers mathematical keys solve can be done within a shorter period than the current system \cite{gerjuoy2005shor}. Therefore, the hardware-based mechanism is one of the alternative approaches to gain over software-based solutions \cite{mohanty2019security}. Hardware-based security mechanisms are depicted in Fig. \ref{FIG:Security_Mechanisms} \cite{shamsoshoara2020survey}.

Dedicated hardware-integrated circuit or processor to store the keys for using a cryptographic function in hardware-based security. For crypto processing and strong authentication hardware security module, which is physical computing device, is used. This module is used for encryption, decryption, storing, and managing digital keys to use in different encryption mechanisms like  Fig. \ref{FIG:Security_Mechanisms} \cite{barker2011transitions}. There are several encryption methodologies such as public-key infrastructure (KPI), attribute based (ABE), Elliptic curve cryptography (ECC), etc. PKI is a system which is based on encryption key pairs and digital certificates. ABE is a type of public-key encryption (PKE) where it uses a set of attributes to generate the secret key. Key-policy attribute-based encryption and ciphertext-policy attribute-based encryption are the types of ABE. ECC can provide similar level of security compared to non-elliptic curve (EC) cryptography with smaller key size. It is also a PKE based on algebraic structure of ECs over finite fields. Another security solution, blockchain, which is decentralized, immutable, distributed, and shared digital ledger to protect the IoV data. It is a peer-to-peer network and instead of central authority, all nodes come to a consensus to verify transactions by miners and store the data in a block to make a chain of blocks \cite{mohanta2020addressing} \cite{singh2020internet}. But man-in-the-middle attacks can breach hardware-based security schemes, and also cloning attack can be happened after stolen the device by an adversary.  Physical unclonable function (PUF) can provide resistance against the attack. PUF as a  security primitive was first introduced by Gassend et al. in \cite{gassend2002silicon}. PUF can produce digital fingerprint (Response against Challenge) on demand and on the fly\cite{yanambaka2018making} \cite{yanambaka2017making}. Among characteristics of PUF, uniqueness, reliability, and randomness can be considered as major \cite{sadhu2021performance} \cite{joshi2017everything} \cite{alladi2020harci}.  

The rest of the paper is organized as follows: Section \ref{SEC:RelatedResearch} discusses the prior research related to the Smart Vehicle and the IoV security and privacy, Section \ref{SEC:Contributions} presents the threat being addressed in the current paper and the proposed solution, Section \ref{SEC:Proposed_Authentication_Framework} presents the proposed Easy-Sec architecture. The results and security analysis on the proposed framework is presented in Section \ref{SEC:Experimental_Results} and the conclusion and future directions are proposed in Section \ref{SEC:Conclusion}.



\section{Related Work}
Many researchers are working to secure the IoV network as it is growing and it depends on many things including human life. The IoV network demands a secure authentication protocol that requires low complexity, less resources, faster authentication. In this section, literature review of existing authentication schemes of the IoV network will be discussed.

In 2017 \cite{guo2017secure}, an authentication scheme for large scale IoV was proposed. It used a certificate that is loaded in AV before registration by a certificate authority. Private-public key would be issued during registration time if the certificate seems valid. This key was used for encryption. ID was not anonymous in this case and also after changing the RSU area, fake certificates could be used to forgery vehicles. A certificate-less scheme \cite{li2019cl} was proposed to avoid certificate storage in TA and AV. To hide the secret key, it used two random values which made the messages unlinkability. It assumed that discrete algorithm was intractable and also it was stated that in future work, the vulnerabilities and limitations of the work would be removed to achieve security and privacy requirements. Trust based scheme was developed \cite{kerrache2018tacashi} in using online social networks of drivers and passengers. Each vehicle calculated trust and RSU decides trust value from these recommendations. Interdevices authentication required Chaotic map based Chebyshev polynomials for computing security keys and also used Advogato trust metric to calculate network trust. This scheme introduced delay as it was required to compute human factor and location related trust and also, trusted third parties like social network platforms, network providers, etc. Ciphertext-policy attribute based encryption (CP-ABE) based system had been proposed in \cite{han2021implementing}. Frequent item sets were built by mining the frequency features using max-miner association rules algorithm. It used keygen algorithm using frequency attributes as input to generate a secret key for decryption. If the same frequency sets were used, then ECU would be able to extract data. It could be exposed as a major vulnerable if an adversary gets the specific attribute set by compromising one ECU. ECC based batch authentication system to authenticate multiple entities was used in both \cite{wu2020batch} and \cite{zhang2020extensible}. In \cite{wu2020batch}, RSU took the help of assistance verification terminals (AVT) depending on the amount of computational power and network traffic. This work was more focused on batch verification rather than security solutions. The stored key in TPM and a random number to avoid side-channel attacks was utilized in \cite{zhang2020extensible}. Each vehicle needed to share a signature when it received the RSU broadcasting certificate. This work did not consider secure communication between RSU and TA. Moreover, there was a chance that vehicle generated PID was in the compromised list. Another ECC based security framework was developed in \cite{thumbur2020efficient}. It aggregated signatures from different vehicles into a single signature to reduce storage at RSU. It might suffer from different security threats like side-channel attacks, DoS attacks. Machine learning is another way to resist threats which were being explored in \cite{sharma2020machine} and \cite{pascale2021cybersecurity}. \cite{sharma2020machine} worked on position-based attacks using six algorithms. This worked on identifying location plausibility and movement plausibility using supervised machine learning to find the pattern and predict the misbehavior of vehicles. But this work was limited to 5 position-based attacks and it did not provide 100\% accuracy. An embedded intrusion detection system in an SoC was used in \cite{pascale2021cybersecurity}. It worked in two steps. In the first step, ten state frames of various parameters were analyzed. In the second step, Bayesian network was used to predict a probabilistic graph. Unfortunately, it showed degraded performance when it was tested on a Free State Attack. 

It was considered that TPM or TA or key generator are part of previously discussed authentication frameworks. These schemes could be affected by side-channel attacks due to storage entities \cite{li2019cl}. To overcome this situation, PUF is incorporated into schemes\cite{sadhu2021performance}. A novel PUF-based \cite{wang2018notsa} on board unit (OBU) was developed to make communication with an external network. This work needed a secure area in OBU. A PUF-based security methods was developed in \cite{labrado2021fortifying} where hash of responses are encrypted for mutual verification. Another authentication scheme was developed in \cite{alladi2020lightweight} where only one CRP was used. This work could be affected by insider attacks and also secure channel between RSUs and the edge server was mandatory. Like \cite{alladi2020lightweight} a single CRP based authentication scheme was designed in \cite{aman2020privacy}. The system could be compromised by its periodic CRP update process.

So far discussed authentication schemes are applicable for centralized systems whereas blockchain is used for decentralized systems. A blockchain based solution to ensure security and privacy of the IoV network in \cite{wazid2022fortifying}. It preserved data from being exposed and modified. A debit-credit based blockchain scheme was proposed in \cite{liu2019novel} to avoid the cold start of new vehicles or users. But this work required protecting signatures and their resources. A high resourceful adversary could compromise the system. An incentive mechanism, where multiple vehicles bid to complete a task using their resources, was developed in \cite{yin2019efficient}. Moreover, the bidding system was avoided in case of emergency conditions where multiple vehicles make a cluster. This scheme took a reactive approach to resist attack instead of a proactive approach. Furthermore, this scheme did not depict inter-vehicle communication for clustering. Vehicles calculated credibility considering the distance between vehicles in the scheme of \cite{yang2018blockchain}. This work used Bayesian Inference Model and uploads ratings to RSUs. This work was suitable against limited resourceful attackers. Also, it did not resist reply attacks, man-in-the-middle attacks, and introduces high overhead. The framework in \cite{gao2019blockchain} used SDN enabled OBU for effective network management and also fog computing for avoiding frequent handovers. However, this work required to be more focused on data transmission trust. The proposed authentication framework \cite{javaid2020scalable} used smart contracts for registering trusted vehicles and blocking malicious ones etc. Both server and vehicle used CRP for trust establishment and a certificate is issued after authentication. This certificate would be used instead of CRP later. The system could be compromised as RSU stores certificates and it is prone to attack. Also, an adversary could gain few responses as this scheme allowed to share responses after getting challenges. Table \ref{TABLE:related_works} shows comparisons of existing authentication schemes.

\label{SEC:RelatedResearch}
\begin{table*}[h!]
	\centering
	\caption{Comparative analysis of related works.}
	\begin{tabular}{p{2cm} p{3cm} p {2cm} p{3cm} p{4cm}}
		\hline
		Citation    & Problem Addressed & Solution Method
		& Advantages & Disadvantages \\ \hline
		Han et al. \cite{han2021implementing}	& Mining frequency features sets to secure communication & CP-ABE & Improved speed & Limited to same attribute set \\ 
		Wu et al. \cite{wu2020batch}	& Reduce the verification delay \& achieve fast message verification & ECC & Batch verification & Need security improvement \\ 
		Zahng et al. \cite{zhang2020extensible}	& Secure communication with limited bandwidth & ECC & Incorporated random number to avoid side-channel attack & Did not provide solution of RSU and TA secure communication\\ 
		Thumbur et al. \cite{thumbur2020efficient}	& Certificateless authentication scheme with reduced verification time and storage space in RSUs & ECC & Avoids pairing operation & Complex calculations with many parameters; Overhead \\ 
		{Sharma et al.} \cite{sharma2020machine}	& Data-centric misbehavior detection & Supervised ML & Shows good accuracy to few types of position based attacks & Limited to 5 position based attacks; It is undefined for novel data\\ 
		Alladi et al. \cite{alladi2020lightweight}	& Combined attestation and authentication for verification of ECU firmware & ECC \& PUF & Involvement of 1 CRP; faster & Need secure memory location in both RSUs, Edge server and Cloud server \\ 
		Aman et al. \cite{aman2020privacy}	& Resisting physical attacks; Improve throughput & PUF & Simple; Single authentication in RSU gateway & Need RSU storage and CRP update \\ 
		Javaid et al. \cite{javaid2020scalable}	& Trust establishment in IoV & PUF \& Blockchain & Improved latency & RSU issues certificates; CRP could be exposed\\ 
		
		\hline
	\end{tabular}
	
	\label{TABLE:related_works}
\end{table*}

\section{Contributions of the Current Paper}
\label{SEC:Contributions}
\subsection{Research Question and Threat Model}
Association for Safe International Road Travel's reported nearly 1.25 million people die yearly, an average of 1,287 deaths a day, in road crashes. Another report in \cite{zadobrischi2021vehicular} also shows that 60\%-70\% road accidents could be avoided which can save many peoples from 1.3 million yearly death and 50 million peoples injury. Human error is the primary reason in 90\% accidents for which the IoV is a potential solution. Though the IoV can increase road safety by monitoring and sending recommendations to the driver on time, it is required to overcome certain difficulties such as communication latency, transmission throughput, power consumption and most importantly security defense systems and privacy protection strategies to develop a complete IoV system  \cite{haque2020lora} \cite{zadobrischi2021vehicular}.

The IoV network requires third party addition, this portability involvement will will make the paradigm open source for attacks. Furthermore, security threats can be increased due to limited bandwidth. Many loss due to the IoV network will not create adverse effect on driver or consumers but also on automobile industry. \cite{bagga2020authentication} states that by modifying or changing data or receiving prank data leading wrong decision, roads can be kidnapped. It is desired to develop and incorporate a robust authentication framework to build an IoV network by resisting security vulnerabilities, protecting privacy \& data integrity and also verifying messages faster.

\subsection{Contributions}
All the existing authentication frameworks are not able to make the IoV network completely secure. A PUF - Based ultra-lightweight authentication scheme has been proposed in this work to improve the security system and resist security vulnerabilities.
\begin{itemize}
	\item PUF is introduced in this work which eliminates storage requirement in AVs, RSUs, RGs completely. It will be used to generate response, the secret information, on the fly which will be required to authenticate both AV and CS.
	\item Designed a new mutual authentication scheme which is ultralight so that simple and less computation is required. 
	\item Computational and communication overhead are reduced significantly.
	\item Authentication time is much less than existing frameworks which makes this scheme suitable for the IoV application.
	\item Introduction of PUF resists physical and node tampering attacks. Also, it makes the system resistant against cloning attack.
	\item Pseudo identity is incorporated in this work to ensure privacy protection.
	\item One time authentication is required. This scheme has simple and faster session update process with minimal overhead.
	\item It is able to resist known attacks. It shows superior performance and security analysis validates the proposed Easy-Sec protocol.
	
\end{itemize}

\subsection{Significance of the Contribution}
The proposed scheme uses PUF which eliminates the use of memory of the IoV. Also, the proposed Easy-Sec will authenticate the IoV faster compared to the existing frameworks. As faster authentication is mandatory for a real-time application like IoV, the proposed framework is better suited to the IoV applications. Moreover, it creates less pressure on communication channel by involving minimal data (bytes) requirements. Furthermore, it can preserve the anonymity of the IoV and preserve security \& privacy by resisting known attacks.

\section{Proposed Authentication Framework}
\label{SEC:Proposed_Authentication_Framework}
This section presents the proposed authentication protocol for the application of the IoV. The proposed authentication framework consists of the following six main system components:
\begin{itemize}
	\item  Drivers/Passengers: Passengers are the primary users in the IoV system who are different individual entities residing in the autonomous vehicle to reach the desired destination. It is required to share real-time information of route such as traffic, which demands strong privacy the in IoV application.
	\item Autonomous Vehicles (AV): AVs which are embedded with ECUs to grab signal, traffic information and computation etc. Moreover, AVs are equipped with PUF for authenticating vehicle in the IoV network. 
	\item Roadside Unit (RSU): RSU is generally described as vehicular communication systems. RSU is a transceiver, resides on the roadside, which sends or receives signal from AVs, pedestrians etc. Furthermore, an RSU operating in a particular location where it is licensed to operate. It supports storage for memory allocation, processor that runs applications for computing capabilities, network capabilities such as 4G/LTE or 5G and also GPS receiver that support secure communications with passing vehicles, other field equipment, and centers. A PUF is integrated into the RSU. It collects messages from AVs and send to RG after combining those.
	\item RSU Gateway (RG): RG provides coverage of a certain location and RSUs of the coverage area connected with the RSU gateway. RSUs are authenticated with RSU gateway for connection. A PUF is integrated to the RG.	An RG combines the messages of its serving RSUs and send to Cloud Server for computation, verification and other tasks.
	\item Cloud Server (CS): Cloud server is the decision maker for establishing communication between entities. It authenticates AVs, pedestrians, RSUs etc before accepting data. Moreover, it is a memory device for storing traffic related information in the IoV application.  
	\item Secure Database (SDB) : SBD is secured storage for keeping CPR set of AVs, RSUs etc which is required for authentication by cloud server. 
\end{itemize}

Each AV wakes up from inactivity mode, it tries to sends \& receives traffic information from the IoV network. Before connecting to the IoV network, it must go through authentication process to identify whether it is legitimate vehicle or user. Each AV must go through two phases for being part of the IoV network. Initially, it is required to be registered and then perform mutual authentication phase.

\subsection{Overview of PUF-based Ultalight Authentication Framework}

\begin{figure}[htbp]
	\centering
	\includegraphics[width=0.75\textwidth]{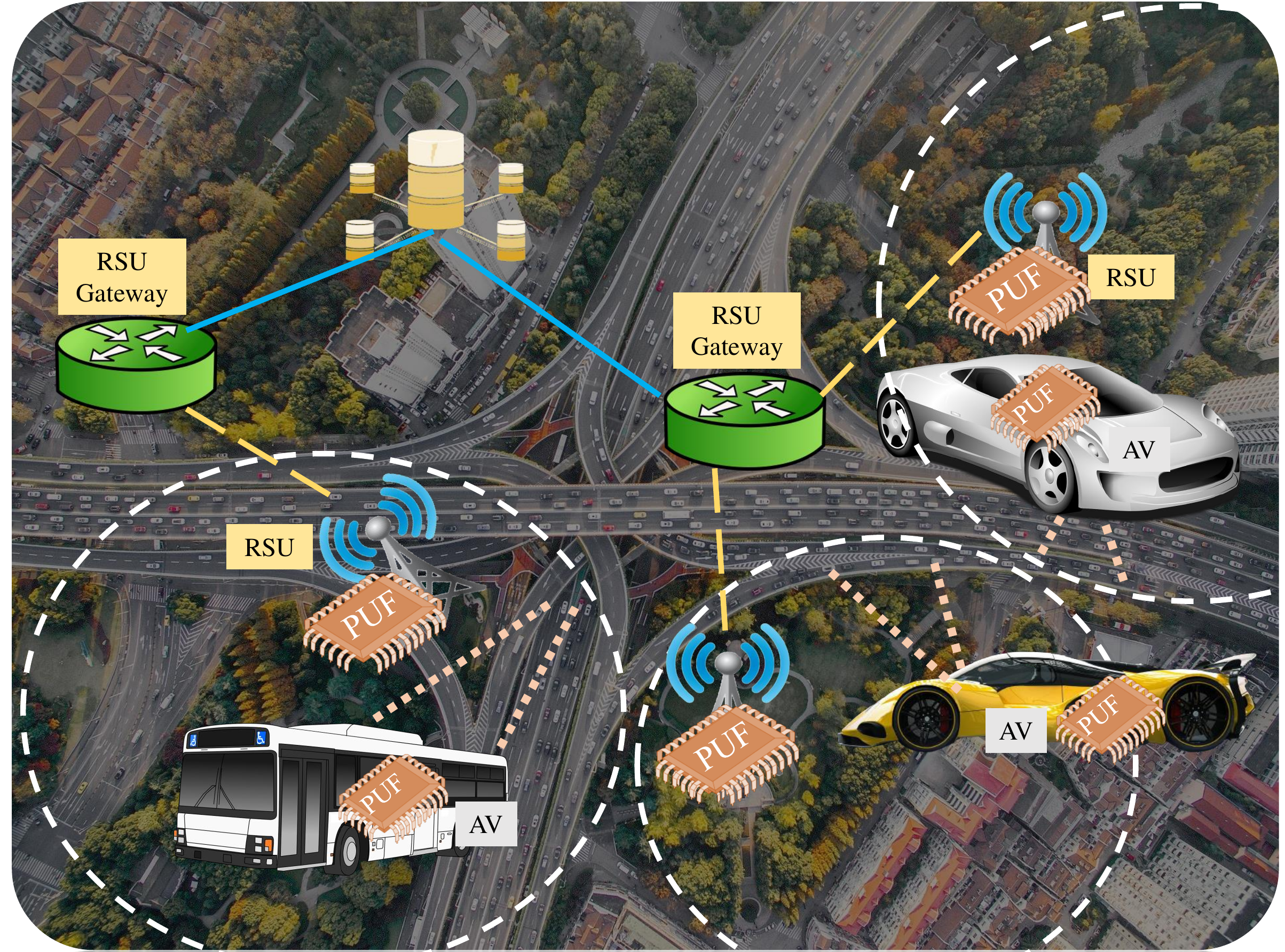}
	\caption{Architecture of the Proposed PUF Based Ultalight Authentication Scheme in the Application of the IoV} 
	\label{FIG:overview}
\end{figure}

Fig. \ref{FIG:overview} shows the proposed PUF based security paradigm. As shown in the figure, AVs are the end devices. Each AV is connected to an RSU and each RSU is connected to an RG. An RG is connected to a number of RSUs and combined coverage area of those RSUs is the serving area of the RG. There are multiple RGs which are connected with a cloud server. When an AV wants to join the IoV network, it will communicate with the CS through the nearest serving RSU, and RG for mutual authentication. After successful authentication, the AV can be part of the network to send and receive required traffic information. For mutual authentication, the CS will ask the AV to generate a response for a selected challenge. If the response matches with the one stored on the SDB, authentication will be successful. A session key will then be shared with the AV for further communication. The AV does not need to authenticate again within the RG coverage area. When the AV changes RG serving area, session keys are required to be updated which avoids further mutual authentication. The symbols which are used in this authentication scheme are presented in Table \ref{TABLE:Acronyms}.   

\subsection{Assumptions}
The following assumptions are considered for secure and successful mutual authentication using proposed framework:
\begin{itemize}
	\item Each AV and RSU, RG are embedded with PUF chips and at the time of AV's registration, AVs which are embedded with PUF will go through CRP generation step in the registration phase through secure channel.
	\item There is a secure connection between CS and SDB. SDB is the only trusted center of CS for storing.
	\item The IDs, PIDs, CRPs of AVs, RSUs and RGs are stored in the SDB of the CS during the registration phase.
	\item No shared keys exists either between AVs and RSU; or RSUs and RGs; or RGs and CS.
	\item RSUs, RGs are already authenticated with CS for communication.
	\item AV does not maintain continuous connection with the RSU, RG and CS, instead it operates in wake-sleep cycles for better energy efficiency. When a user needs to go to a place, it wakes up and establish secure session by authenticating mutually. 
	\item It is considered that the PUFs which are used in this scheme are noise resistant and perform in the same way in every environment and life span. Like \cite{lu2018cmos} and \cite{chuang2019physically}, there are many noise resistant PUFs have been developed in recent times to ensure the reliability which can deal with environmental issues, voltage fluctuations, wide temperature ranges, pressure and humidity etc.
\end{itemize}

\subsection{PUF-based AV Registration Phase}

\begin{figure}[htbp]
	\centering
	\includegraphics[width=.5\textwidth]{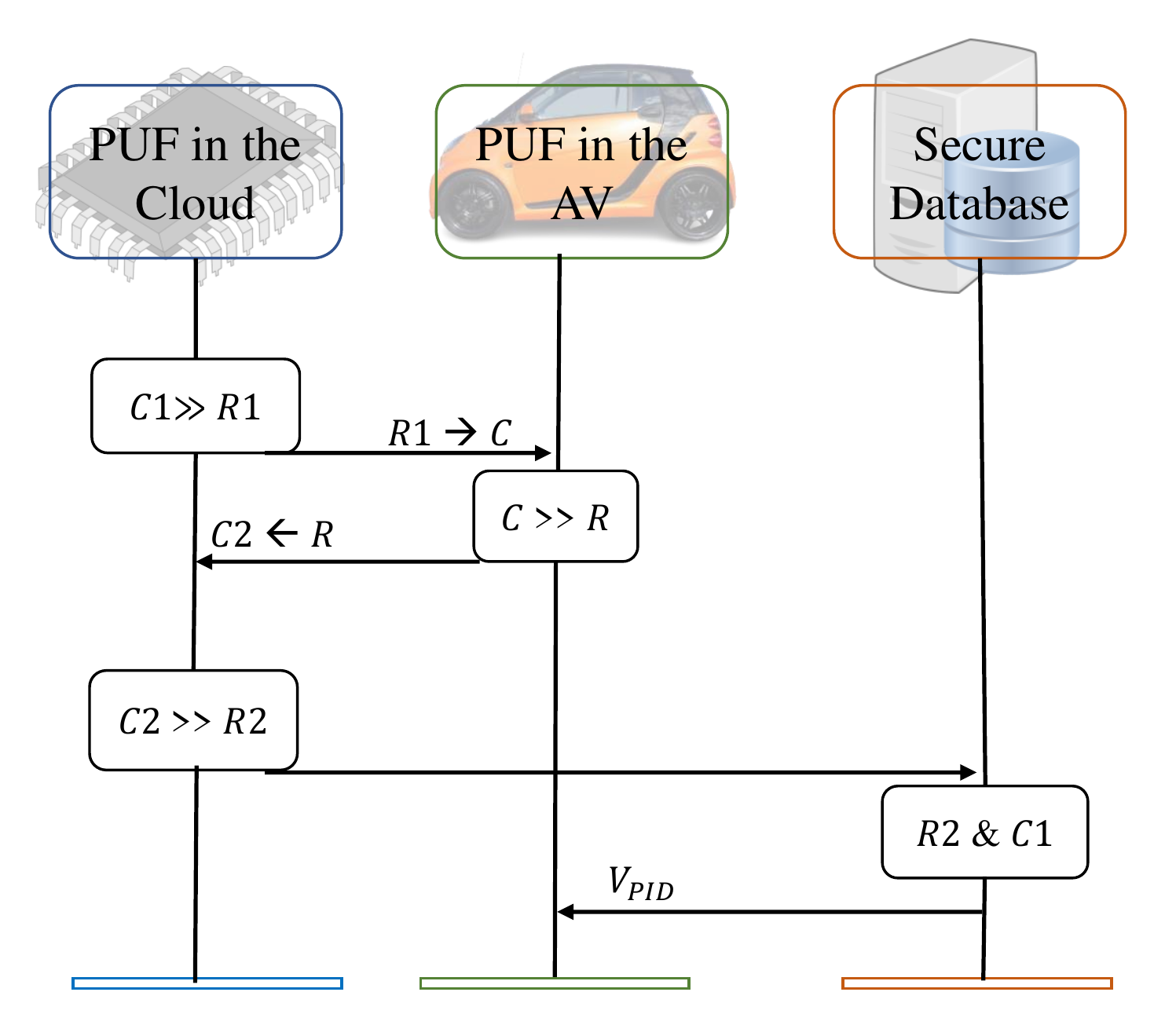}
	\caption{Easy-Sec Vehicle Registration}
	\label{FIG:registration}
\end{figure}

When a client purchased a new AV and want to enroll in the IoV network, it is first required to go through registration phase. Every AV which needs to be registered in the IoV network must have a integrated PUF module. Registration process needs to be performed through a secure channel and secure environment and the PUF module in the vehicle satisfies the required characteristics of PUF which is discussed in the section \ref{SEC:Introduction}. Fig. \ref{FIG:registration} shows the registration process of the AV for the proposed authentication scheme. Both AV and CS have an integrated PUF module.

A challenge ``	$C1$'', selected will be given to the PUF at CS and a response, ``	$R1$'' is generated. This response ``	$R1$'' is sent to the AV which is given as a challenge for the integrated PUF. The response, ``	$R$'' generated at the PUF in AV is sent back to the CS. This ``	$R$'' becomes the challenge for the PUF at the AV and generates a response, ``	$R2$''. 
The response $R2$ is stored in the SDB along with the initial challenge. To maintain the privacy of the AV, a pseudo random identity is sent to the AV for future communications.  The control flow of the registration process of a AV is represented by Fig. \ref{FIG:registration}. Here, Challenge-Response relation is depicted by $>>$ sign. For instance, 	$C2$$>>$	$R2$ represents 	$C2$ as Challenge (input) and 	$R2$ as Response (output). On the other hand, $\rightarrow$ sign indicates that transmitted output will act as input to the receiver. For example, 	$R$ $\rightarrow$ 	$C2$ shows 	$R$ will be transmitted and it will be input 	$C2$ of the CS. 


Because of unique process variation of chips, each CRP of a PUF will act as an unique fingerprint of the vehicle. For storing a CRP, it is required to perform the CRP generation  multiple times to check the reliability of the module. This process is repeated for other challenges and corresponding hash values are computed; then the pair is stored in the SDB for the PUF. It is noted that no CRP is stored in the AV in this process. AV will only generate response when a challenge is provided to the AV. The AV registration flow is shown in Algorithm \ref{ALG:AV_Registration}. AV is introduced into the network after the successful registration.

\begin{algorithm}
	\SetAlgoLined
	\quad CS: \\
	\quad \quad \quad 	$C1$$>>$	$R1$ \Comment{In the CS, $C1$ will act as a challenge and will generate response $R1$}\\ 
	\quad CS $\to$ AV \{ 	$R1$, i.e. 	$C$ \} \Comment{The CS will send $R1$ to AV. The $R1$ is same as challenge $C$ in the AV}\\
	\quad \quad \quad 	$C$ $>>$ 	$R$ \\
	\quad AV $\to$ CS \{	$R$, i.e. 	$C2$ \} \\
	\quad \quad \quad 	$C2$ $>>$ 	$R2$ \\
	\quad CS $\to$ SDB \\
	\quad \quad \quad $\in$ \{$C1, R2$\} \Comment{Storing CRP in the SDB}\\
	\quad SDB $\to$ AV \\
	\quad \quad \quad $\in$ \{$V_{PID}$\} \Comment{After storing CRPs, a $V_{PID}$ will be assigned to the AV and will be stored to the SDB}\\	
	\caption{\textbf{AV Registration}}
	\label{ALG:AV_Registration}
\end{algorithm}

\subsection{Easy-Sec: Proposed Authentication Scheme}

\begin{figure}[htbp]
	\centering
	\includegraphics[width=0.55\textwidth]{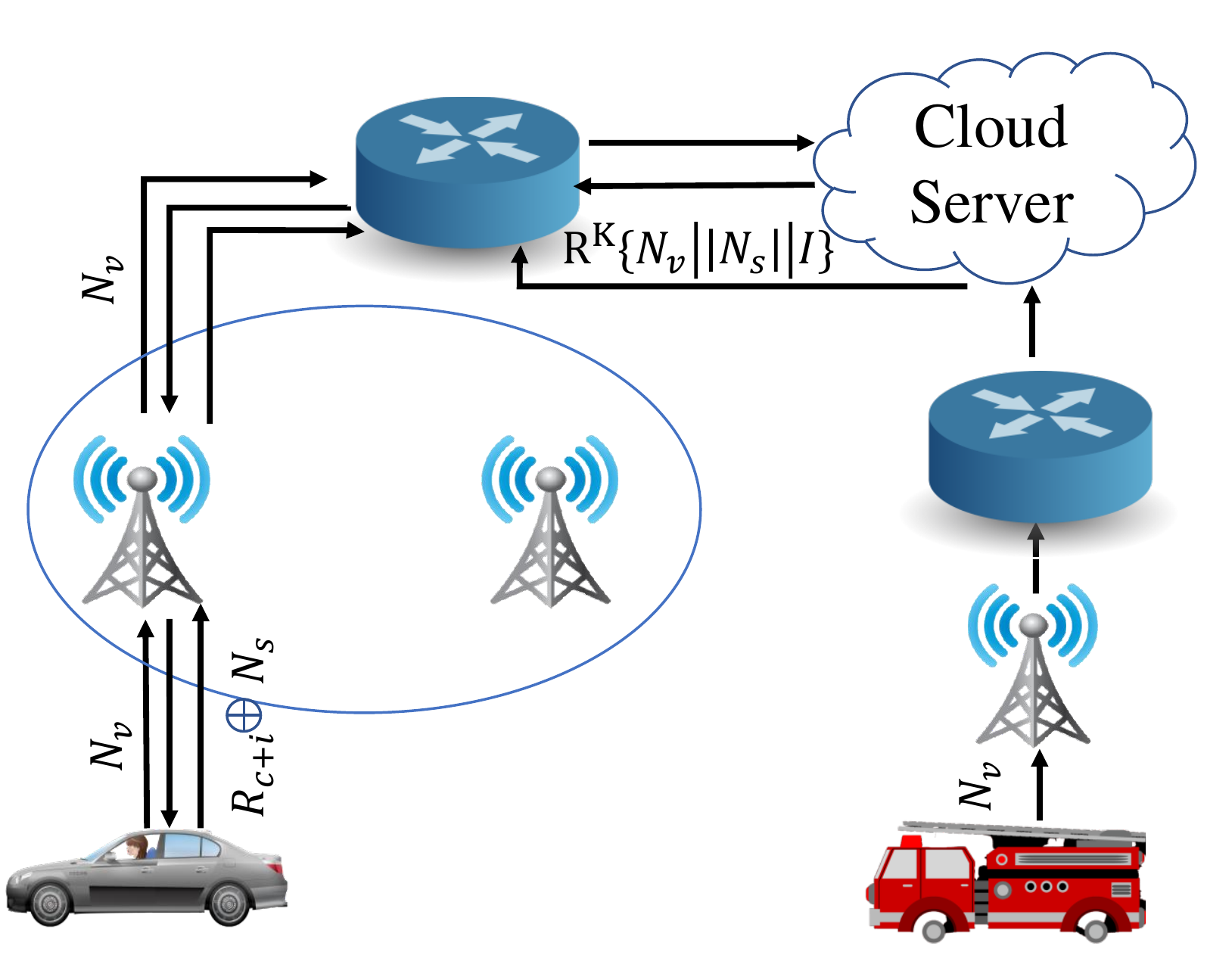}
	\caption{Easy-Sec Authentication Protocol}
	\label{FIG:Authentication}
\end{figure}

\begin{figure}[htbp]
	\centering
	\includegraphics[width=0.75\textwidth]{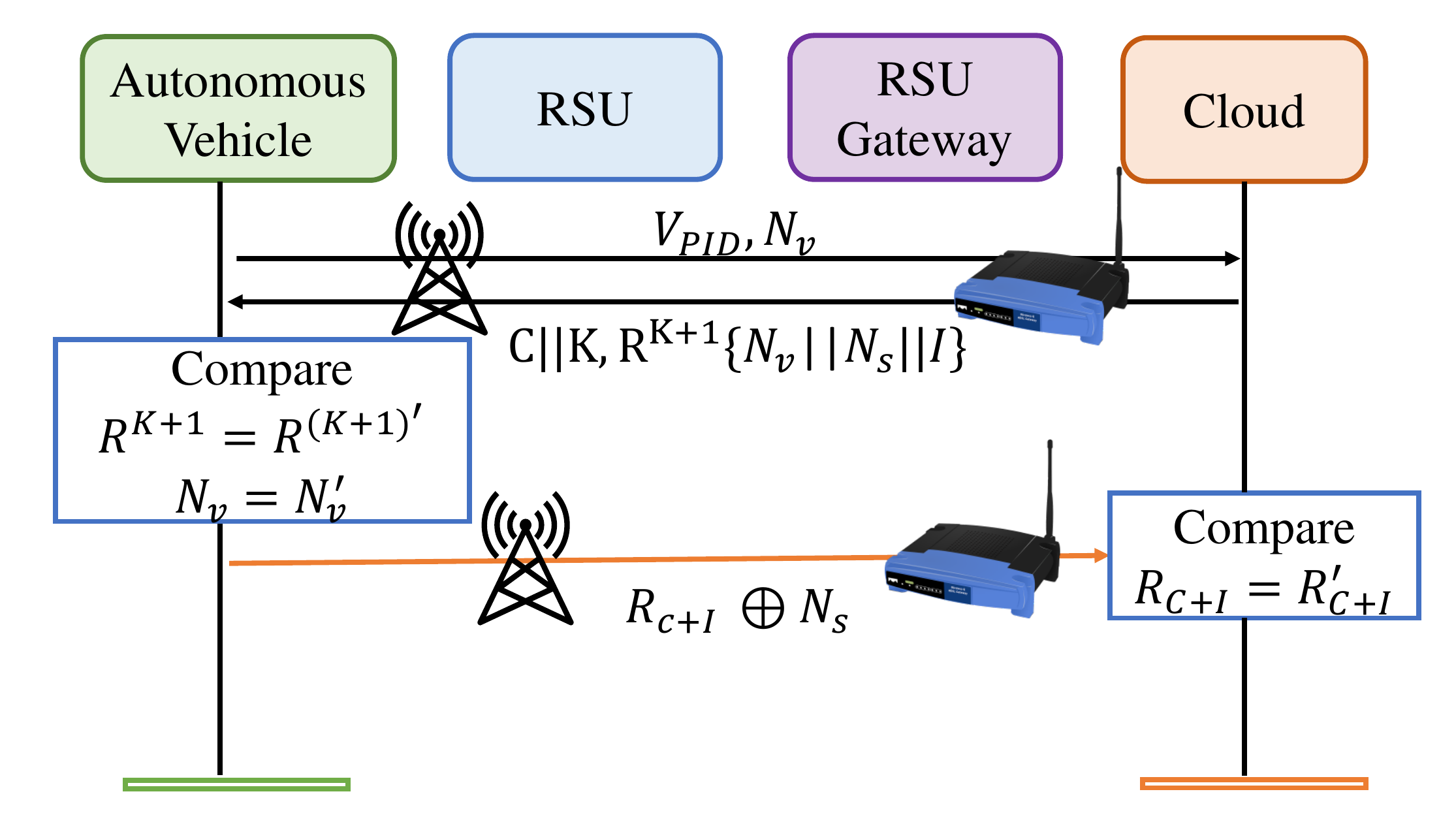}
	\caption{Easy-Sec Authentication Verification}
	\label{FIG:verification}
\end{figure}

\begin{figure}[htbp]
	\centering
	\includegraphics[width=0.75\textwidth]{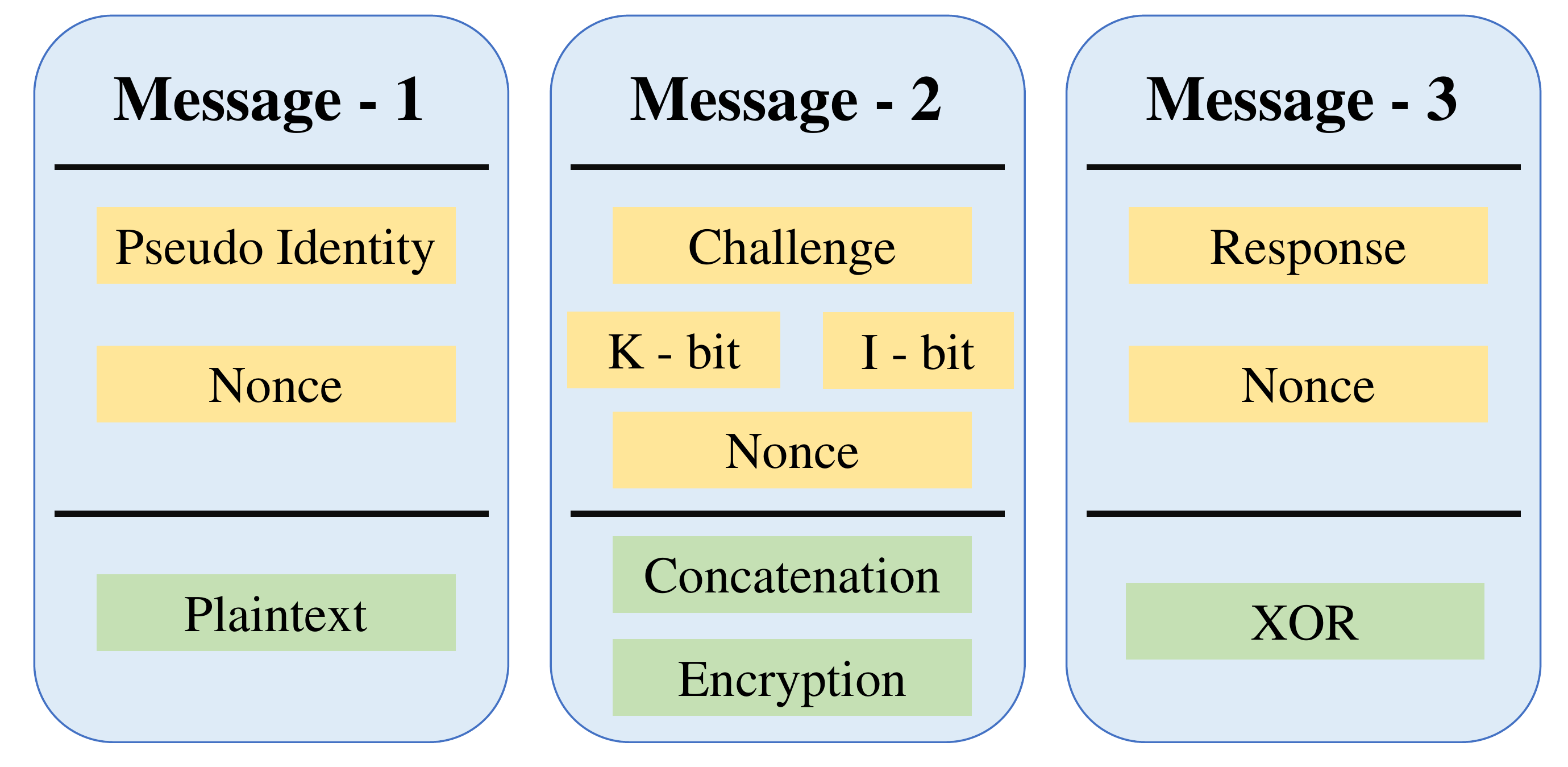}
	\caption{Message Structure in Easy-Sec}
	\label{FIG:Overhead_protocol}
\end{figure}

The proposed authentication scheme is presented in Fig. \ref{FIG:Authentication}. It shows that both server and vehicle authenticate each other communicating through the RSU and RG. The flow of verification using the proposed framework is represented by Fig. \ref{FIG:verification}. For brevity, the message flow of RSUs and RGs are not shown in the Fig. \ref{FIG:verification} as RSUs pass messages from AVs and RGs pass message from RSUs. The authentication process can be divided into 3 phases. Authentication flow is shown in Algorithm \ref{ALG:Authentication_Process} and Fig. \ref{FIG:Overhead_protocol} shows the message blocks of the proposed authentication scheme.

 As stated earlier, it is assumed that PUF based RSU and RG are already been authenticated using CRP communication. 

\begin{enumerate}
	\item \textbf{Authentication session initiation from AV:} When an AV wakes up and initiates a connection to the network, it sends $V_{PID}$ and a nonce $N_{v}$ to RSU for starting a session. RSU passes this packet to the RG and RG to CS. CS will identify the AV using $V_{PID}$ and $ID$ from its stored SDB. The data flow from AV to RSU to RG to CS is illustrated as \textbf{Phase-1} of \textbf{Authentication process}. 
	
	\item \textbf{Verification of CRP from CS:} During this phase, AV will fetch the challenge  and random nonce from CS. CS will send $R^{K+1}$ encrypted nonce $N_{v}$, its own generated nonce $N_{s}$ and a random number to find the challenge. When AV gets the data from CS through RG and RSU, it generates the response using the challenge $C$ and find out $K+1$ bit to decrypt the data to find out information. After decryption, it checks whether whether the nonce it matches or not. If AV finds out the same nonce in the decrypted data, then it verifies the CS. The server verification is presented in \textbf{Phase-2} of \textbf{Authentication process}.
	
	\item \textbf{AV Authentication Confirmation:} In the phase-3, AV generates response of challenge $C+I$ and shares with nonce $N_{s}$ to CS via RSU and RG after doing XOR operation. CS verifies $R_{(C+I)}$ and $N_{s}$ after performing reverse XOR operation. If all these values match, then CS will identify AV as authenticated entity and establish a session key for further communication with RSUs within the serving region of RG. The entire process of this phase is demonstrated in \textbf{Phase-3} of \textbf{Authentication process}.
\end{enumerate}

\begin{algorithm}
	\SetAlgoLined
	\textbf{Phase-1: Authentication session initiation from AV} \\
	AV $\to$ RSU \{$V_{PID}$ $||$ $N_{v}$\} \Comment{The AV initiates authentication by sending concatenation of $PID$ of AV and a random nonce }\\
	RSU $\to$ RG \{$V_{PID}$ $||$ $N_{v}$\}\\
	RG $\to$ CS \{$V_{PID}$ $||$ $N_{v}$\}\\
	\eIf{$V_{PID}$ == $V_{PID^{'}}$}{  
		Continue \Comment{The CS will the validity of the AV $PID$}\\
	}{
		Invalid Client
	}

	\textbf{Phase-2: Verification of CRP from CS} \\
	CS: \\
	\quad $F2_{AV}$ = {$($ $N_{v}$ $||$ $N_{s}$ $||$ $I$  $)_{R^{K+1}}$} \Comment{The CS will encrypt using $(K+1)$ bit of response $R$ for challenge $C$}\\
	CS $\to$ RG \{$C$ $||$ $K$, $F2_{AV}$\} \Comment{The CS will send concatenation of challenge $C$ and $K$ along with encrypted result }\\
	RG $\to$ RSU \{$C$ $||$ $K$, $F2_{AV}$\}\\
	RSU $\to$ AV \{$C$ $||$ $K$, $F2_{AV}$\}\\
	\quad $C$ $\longrightarrow$  $R^{K+1}$ \Comment{The AV will generate $(K+1)$ bit response using challenge $C$}\\
	\quad \{$N_{v}$, $N_{s}$, I = ${R^{K+1}}$ $($ $F2_{AV}$ $)$ \} \Comment{The AV will decrypt the encrypted result to find $N_v$, $N_s$, and $I$}\\
	\eIf{$N_{v}$ == $N_{v}^{'}$}{ 
		Valid Server \Comment{Authenticity of the CS will checked by verifying $N_v$}\\
	}{
		Invalid Server
	}

	\textbf{Phase-3: AV Authentication Confirmation} \\
	AV: \\
	\quad $C$$+$$I$ $\longrightarrow$  $R_{C+I}$ \Comment{Response will be generated using challenge $(C+I)$ in the AV}\\
	\quad $F3_{AV}$ = $F_{nl}$($R_{C+I}$ $\oplus$ $N_{s}$)  \Comment{XOR result of generated response and the CS shared nonce will be sent to the CS through the IoV network}\\
	AV $\to$ RSU \{$F3_{AV}$\} \\
	RSU $\to$ RG \{$F3_{AV}$\} \\
	RG $\to$ CS \{$F3_{AV}$\} \\

	\quad \{$R_{C+I}$ $\oplus$ $N_{s}$\} = $F3_{AV}$ \\
	\quad $Z$ = $R_{C+I}$ $\oplus$ $N_{s}$ \\
	\eIf{$Z$ == $Z^{'}$}{ 
		Authenticated \& Establish Session Key \Comment{The CS will verify the XOR result of response and nonce}\\
	}{
		Authentication Failed
	}	
	\caption{\textbf{Secure Authentication Process}}
	\label{ALG:Authentication_Process}
\end{algorithm}

\subsection{Session Key Update}
\begin{figure}[htbp]
	\centering
	\includegraphics[width=0.65\textwidth]{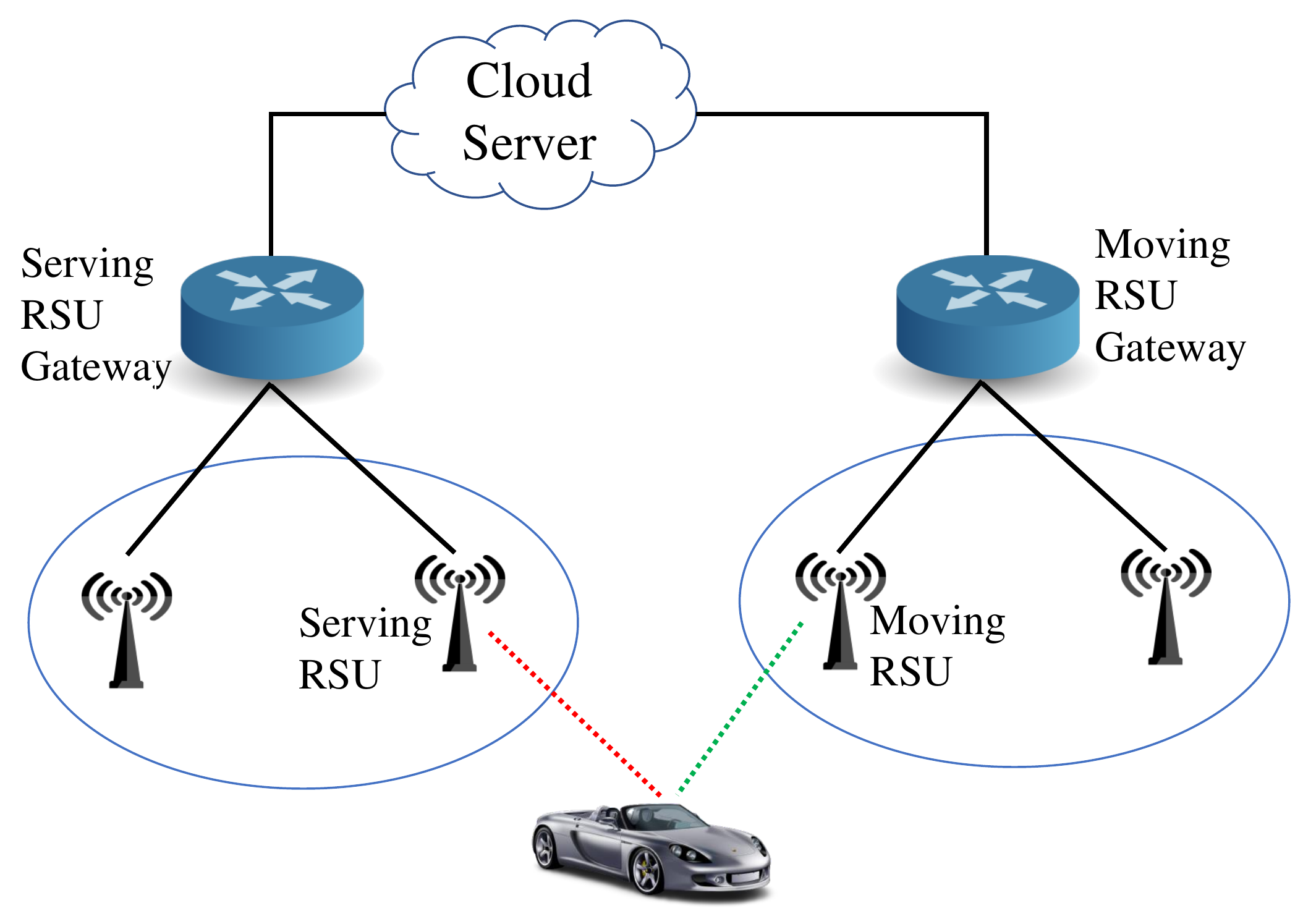}
	\caption{Session Key Update Process}
	\label{FIG:Session}
\end{figure}

When an AV reaches the end of the coverage area of the serving RG as Fig. \ref{FIG:Session}, then it is required to update session key to connect with a RSU of the moving RG. Session key update process excludes future authentication requirement.

\begin{figure}[htbp]
 	\centering
 	\includegraphics[width=0.75\textwidth]{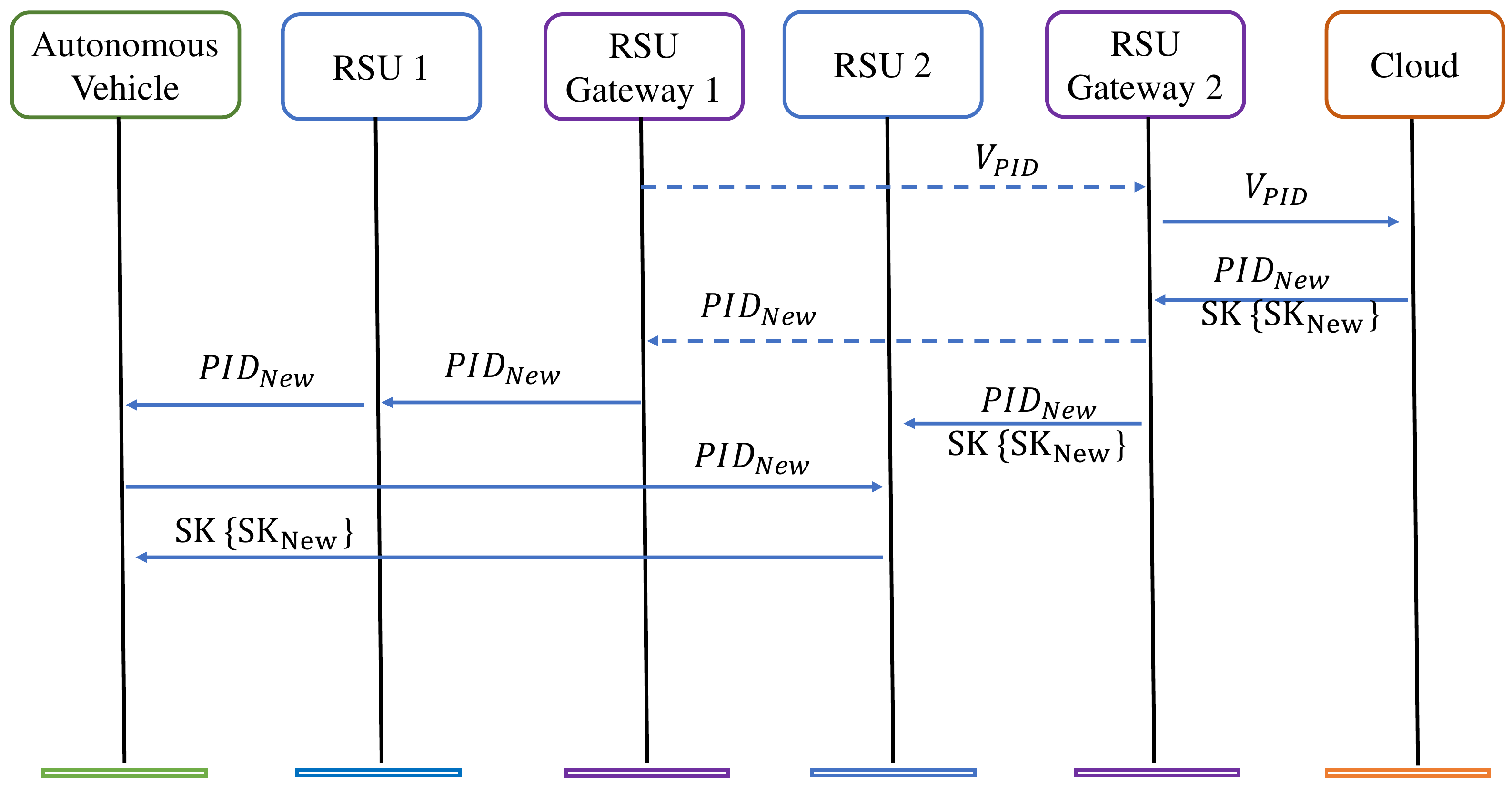}
 	\caption{Procedure of Session Key Update}
 	\label{FIG:Session_key}
 \end{figure}

 Red signal indicates that the signal strength of the serving RSU is becoming weak and green signal indicates that AV is getting strong signal from the moving RSU. The flow of session key update is represented in the Fig. \ref{FIG:Session_key}. The session key update process can be divided into 2 phases.
 
\begin{enumerate}
	\item \textbf{New Session Key Generation:} When RG finds that an AV is about to cross it serving area, then it will initiate session key update process. It will send $V_{PID}$ to the new RG which is RG2 in the figure to. RG2 will communicate with CS for new session key and new $PID$. CS will check the existence of the $PID$ and will generate $PID_{New}$ and session key $SK_{New}$. Then CS will share $PID_{New}$ and encrypted $SK_{New}$ to RG2. After getting information from CS, RG2 will share $PID_{New}$ to RG1 and both $PID_{New}$ \& encrypted $SK_{New}$ to the new RSU which is RSU2 in the figure. AV will get $PID_{New}$ via RSU1 and this $PID_{New}$ will be valid for 1 minute so that an adversary can not attempt to impersonate the AV. These process is presented in \textbf{Phase-1} of \textbf{Session key update process}.
	
	\item \textbf{New Session Key Receiving by AV:} In this phase, AV will initiate communication with RSU2 by sending $PID_{New}$$+1$. Then, RSU2 will share the encrypted $SK_{New}$ with AV. Using current session key, AV will decrypt the message and find out the new session key for making communication in the new RG serving area. The demonstrated process is shown in \textbf{Phase-2} of \textbf{Session key update process}.
\end{enumerate} 

\begin{algorithm}
	\SetAlgoLined
	\textbf{Phase-1: New Session Key Generation} \\
	RG1 $\to$ RG2 \{$V_{PID}$\} \Comment{RG1 initiates session key update process}\\
	RG2 $\to$ CS \{$V_{PID}$\}\\
	CS $\to$ RG2 \{$PID_{New}$, $ ($ $SK_{New}$ $)_{SK}$\} \Comment{The CS shares a new $PID$ of the AV and encrypts a session key using current session key}\\
	RG2 $\to$ RG1 \{$PID_{New}$\} \Comment{RG2 shares the new $PID$ of the AV to RG1}\\
	RG2 $\to$ RSU2 \{$PID_{New}$, $ ($ $SK_{New}$ $)_{SK}$\}\\
	
	\textbf{Phase-2: New Session Key Receiving by AV} \\
	AV $\to$ RSU2 \{$PID_{New}+1$\} \Comment{The AV requests for the session key}\\
	RSU2 $\to$ AV \{ $ ($ $SK_{New}$ $)_{SK}$\} \Comment{RSU shares the encrypted session key}\\
	AV: \\
	\quad \{$SK_{New}$ = ${SK}$ $($ $SK_{New}$ $)$ \} \Comment{The AV decrypts the session key to get the new session key} \\	
	\caption{\textbf{Session Key Update Process}}
\end{algorithm}

\section{Experimental Results and Security Analysis}
\label{SEC:Experimental_Results}
\subsection{Experimental Setup}
\begin{figure}[htbp]
	\centering
	\includegraphics[width=0.75\textwidth ]{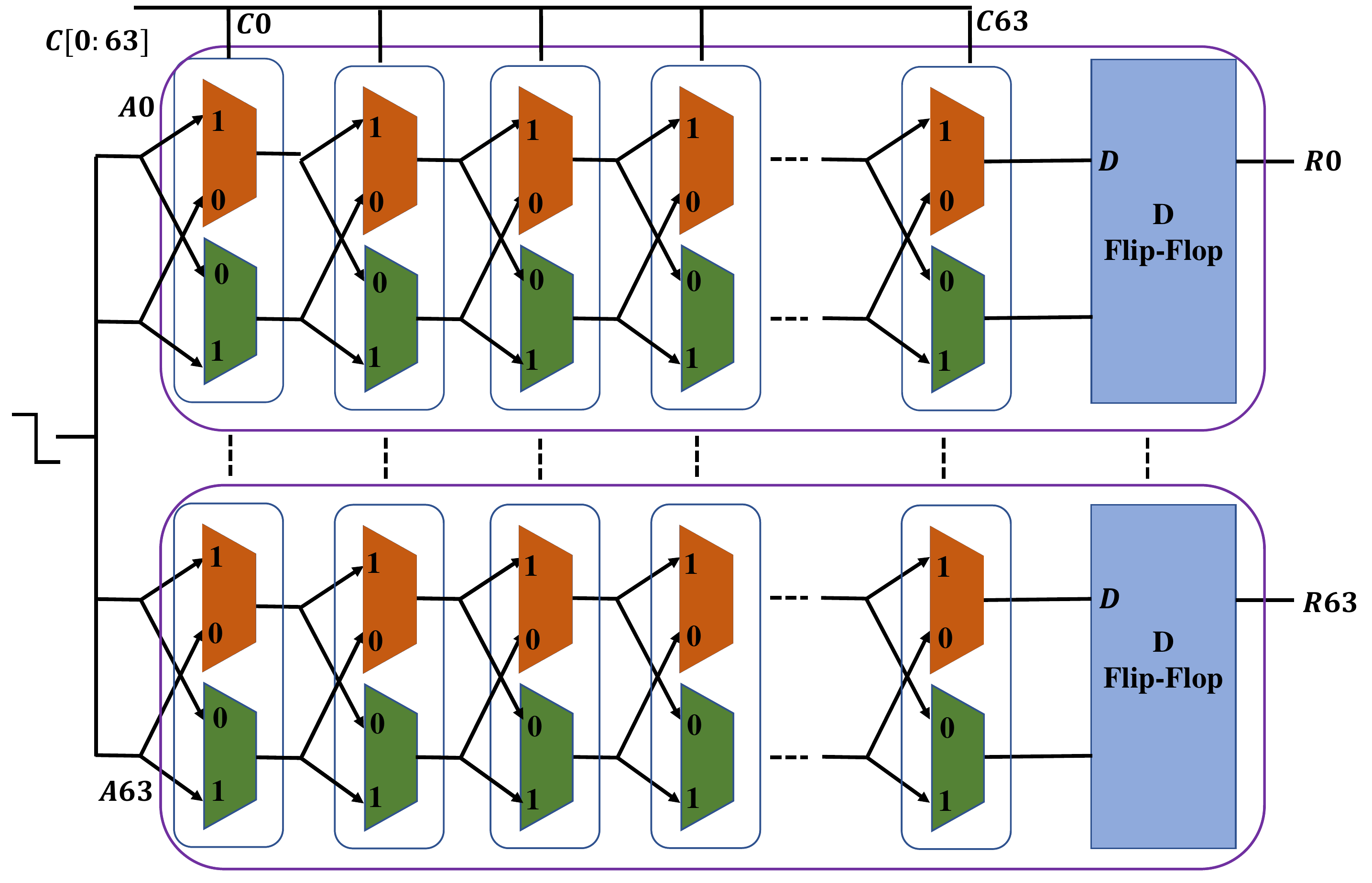}
	\caption{Architecture of 64-bit Arbiter PUF}
	\label{FIG:Arbiter}
\end{figure}

There are many PUF architectures which can generate CRP following the required characteristics. In this work, 64 bit arbiter PUF was used among various PUFs. PYNQ™ Z2 FPGA which is based on Xilinx Zynq C7Z020 SoC was used for PUF implementation. Also, Xilinx BASYS3 FPGA was used to build the PUF. 64 bit arbiter PUF architecture is presented in the Fig. \ref{FIG:Arbiter}. An arbiter PUF is a delay-based PUF that generates a response based on two delay-lines time differences. Each box from $A0$ to $A63$ represents a unit of two delay-lines to generate a bit. Inside each box, a line of orange and green color boxes (2*1 multiplexers) indicates two separate lines for traversing signals. Each bit challenge will act as a selection bit of a pair of multiplexers. Challenge $C0$ will be in the first box from $A0$ to $A63$, $C1$ in the second box from $A0$ to $A63$, and so on. When a signal will be provided in the PUF, the signal will flow through the multiplexers accordingly to the selection-bit from the challenge. For example, if the $C0$ is 0, then the signal will go to the 0 of the multiplexer, if the $C1$ is 1 then the signal will follow the line of 1 of the next pair of multiplexers (between orange and green). In the end, each box of the multiplexers is connected to a D flip-flop. If the input $D$ of the flip-flop gets the signal faster, then the output of the flip-flip means the response bit will be 1. In this way, from a 64-bit challenge (from $C0$ to $C63$) 64-bit response (from $R0$ to $R63$) has been achieved using multiplexer pairs (from $A0$ to $A63$).

\begin{figure}[htbp]
	\centering
	\includegraphics[width=0.85\textwidth ]{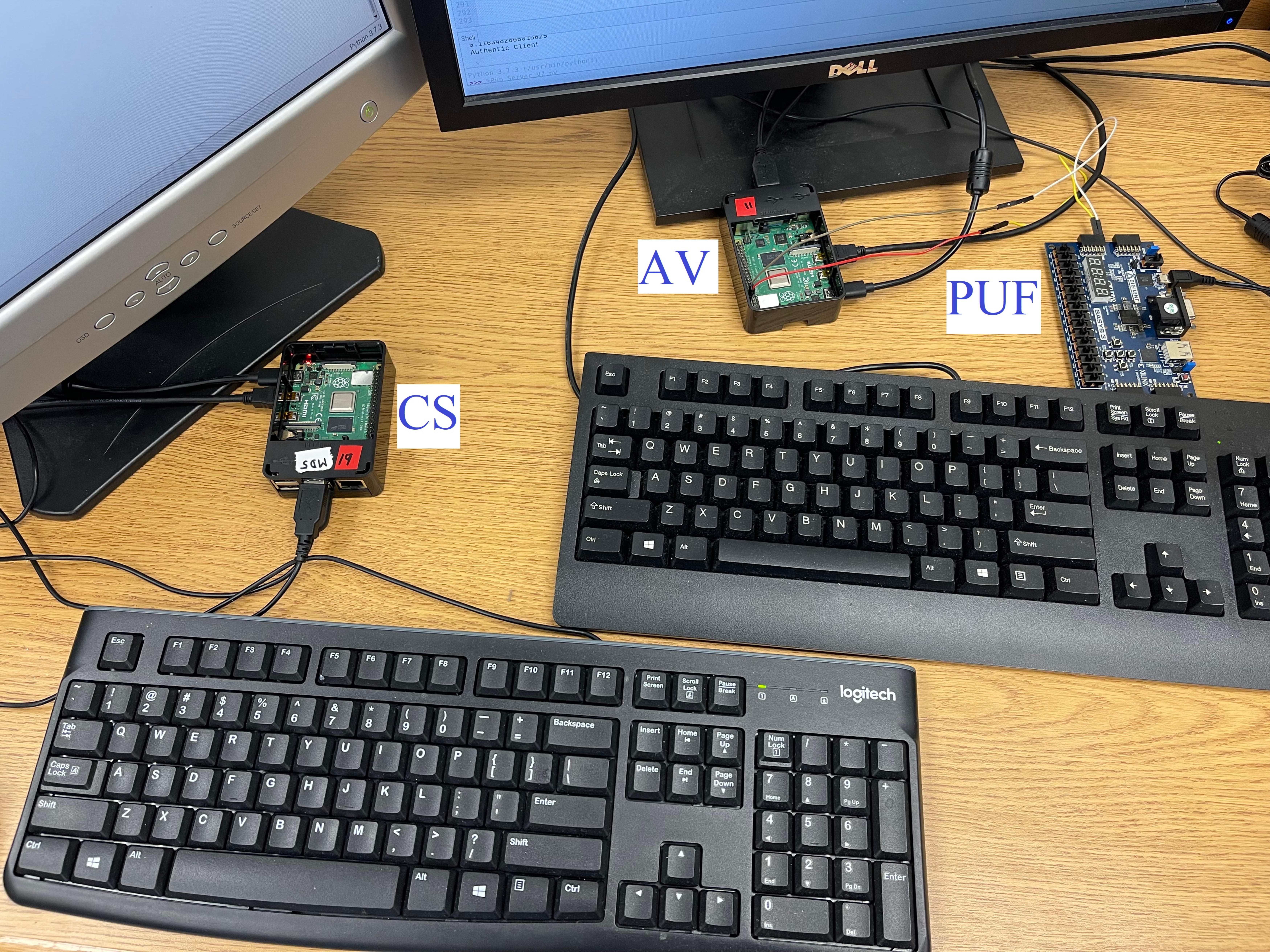}
	\caption{Experimental Setup of the proposed protocol: Easy-Sec}
	\label{FIG:Architecture}
\end{figure}

Raspberry Pi 4 B+, PYNQ Z2 FPGA, and BASYS3 FPGA were used for implementing the work. The experimental setup of the proposed Easy-Sec is shown in the Fig. \ref{FIG:Architecture}. In the figure, it is shown that FPGA is being used as PUF of AV and Raspberry Pi is being worked as AV which are connected together. On the other hand, another Raspberry Pi is working as CS which has a secure database. When authentication request is being sent to CS by AV, CS transmits message which is encrypted using $R^{K+1}$ and in response, AV shares $R_{C+I}$ with nonce of CS. Using the full process, both CS and AV verification will be done so that each can identify whether other end is legitimate or not.

\subsection{Results}
\begin{figure}[hbt!] 
	\centering
	\begin{subfigure}{0.33\linewidth}
		\centering 
		\includegraphics[width=0.99\linewidth]{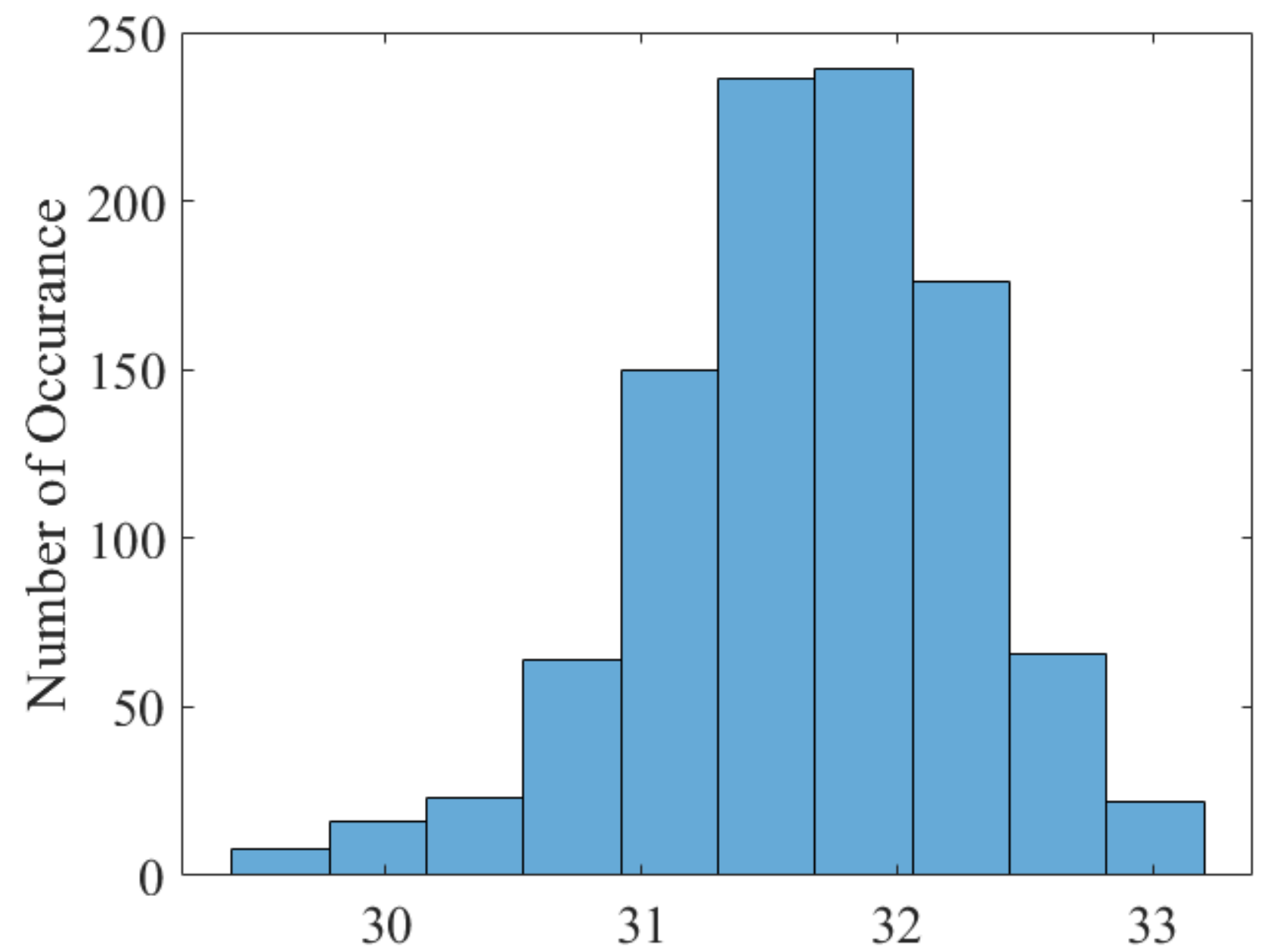}
		\caption{Uniqueness}
	\end{subfigure}
\begin{subfigure}{0.33\linewidth}
	\centering
	\includegraphics[width=0.99\linewidth]{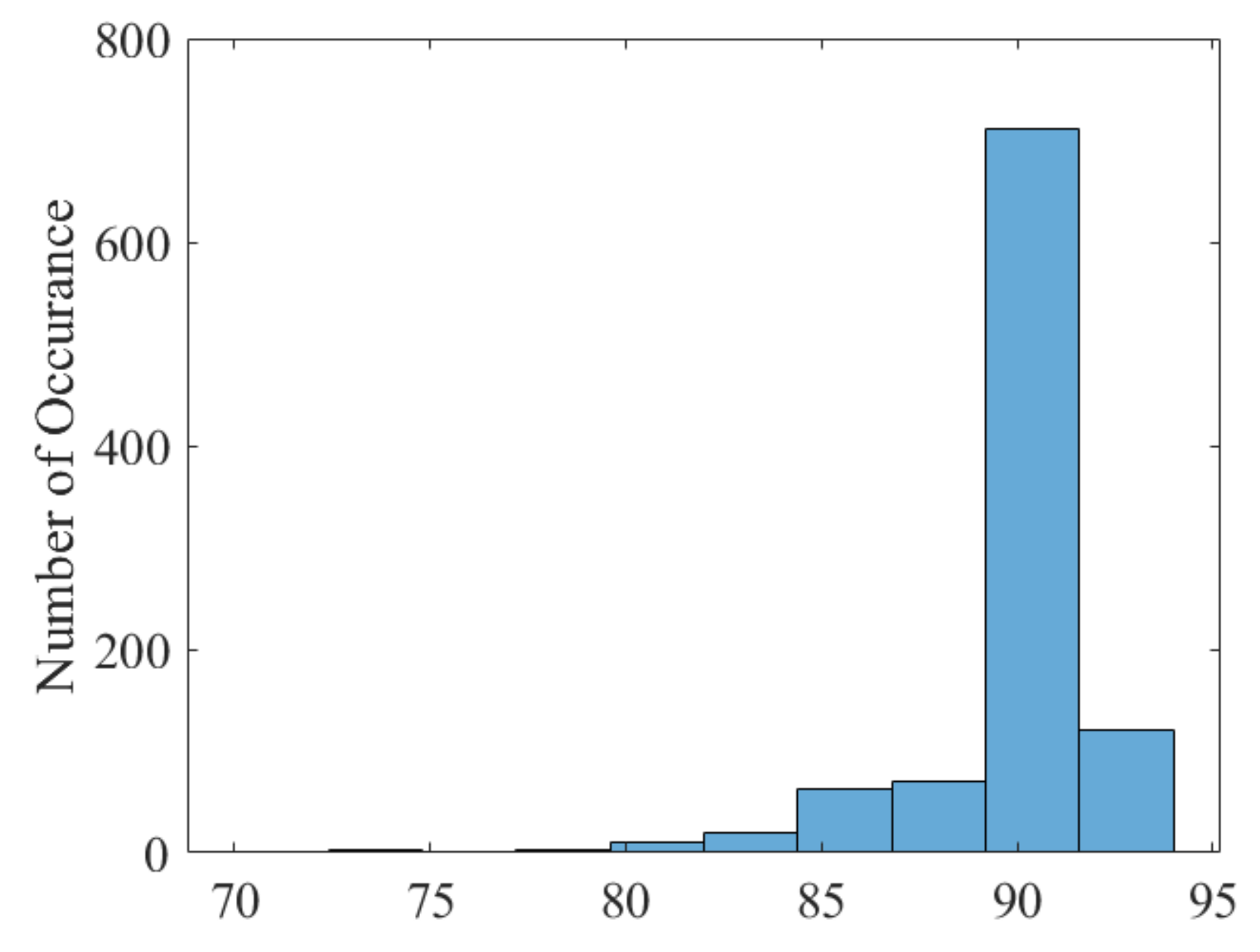}
	\caption{Randomness}
\end{subfigure}%
\begin{subfigure}{0.33\linewidth}
	\centering
	\includegraphics[width=0.99\linewidth]{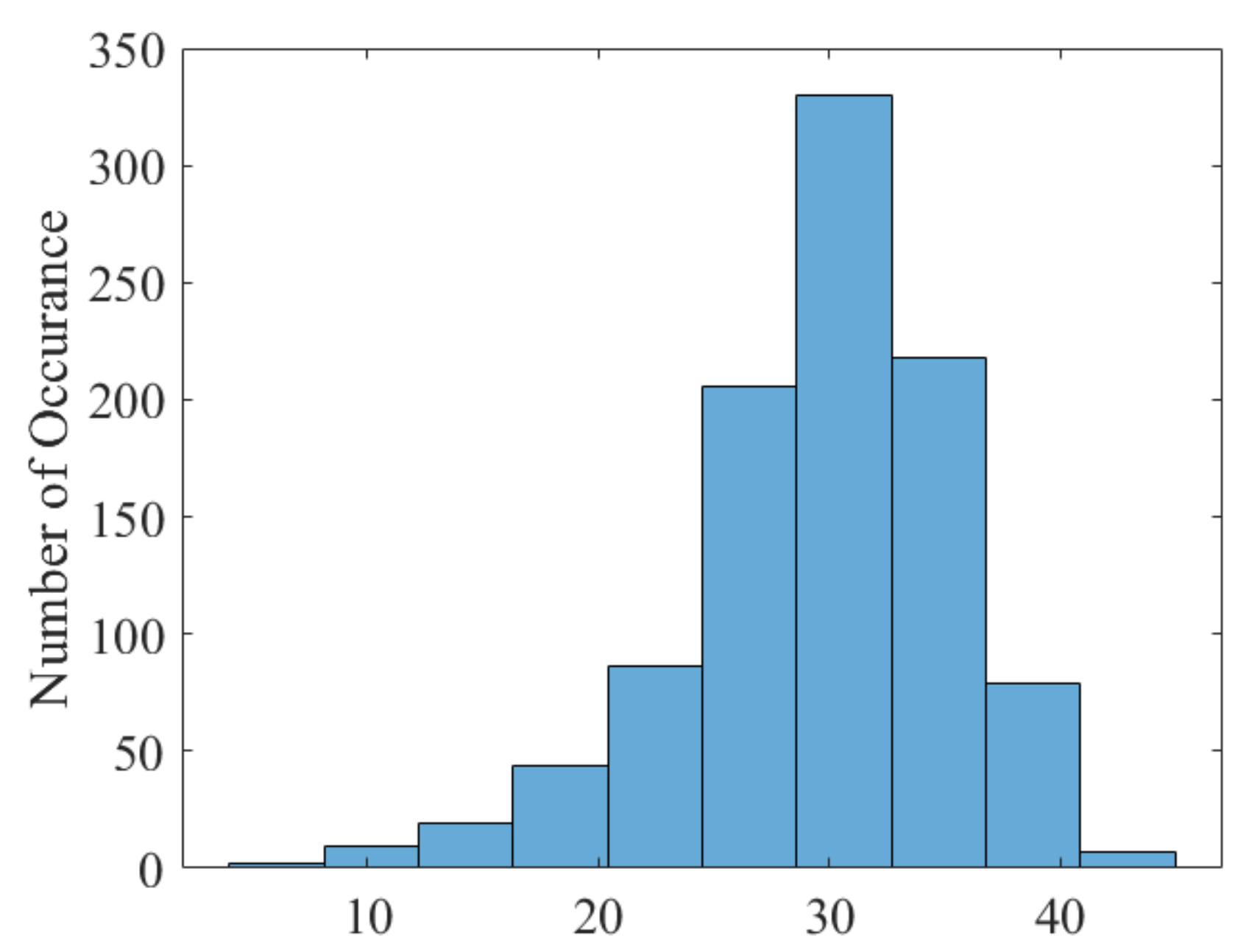}
	\caption{Inter-HD}
\end{subfigure}%
	\caption{Figures of Merit of 64-bit PUF}
\label{FIG:Characteristics_PUF}
\end{figure}

In this section, performance of the proposed authentication protocol will be analyzed. First, the performance of PUF will be presented. Fig. \ref{FIG:Characteristics_PUF} shows the characteristics of the 64-bit PUF that were used in the proposed authentication protocol. The performance was measured using 1000 CRPs. It was found that the PUF showed 31.65\% Uniqueness, 89.69\% Randomness, and 29.5\% inter-HD which can be identified as good performance. Moreover, the PUF showed 100\% reliability. Furthermore, the reliability of PUF was measured at 15$^{\circ}$F intervals from 30$^{\circ}$F to 150$^{\circ}$F temperature scale. The PUF showed robustness as each temperature 100\% reliability  was found. 

\begin{figure}[htbp]
	\centering
	\includegraphics[width=0.55\textwidth]{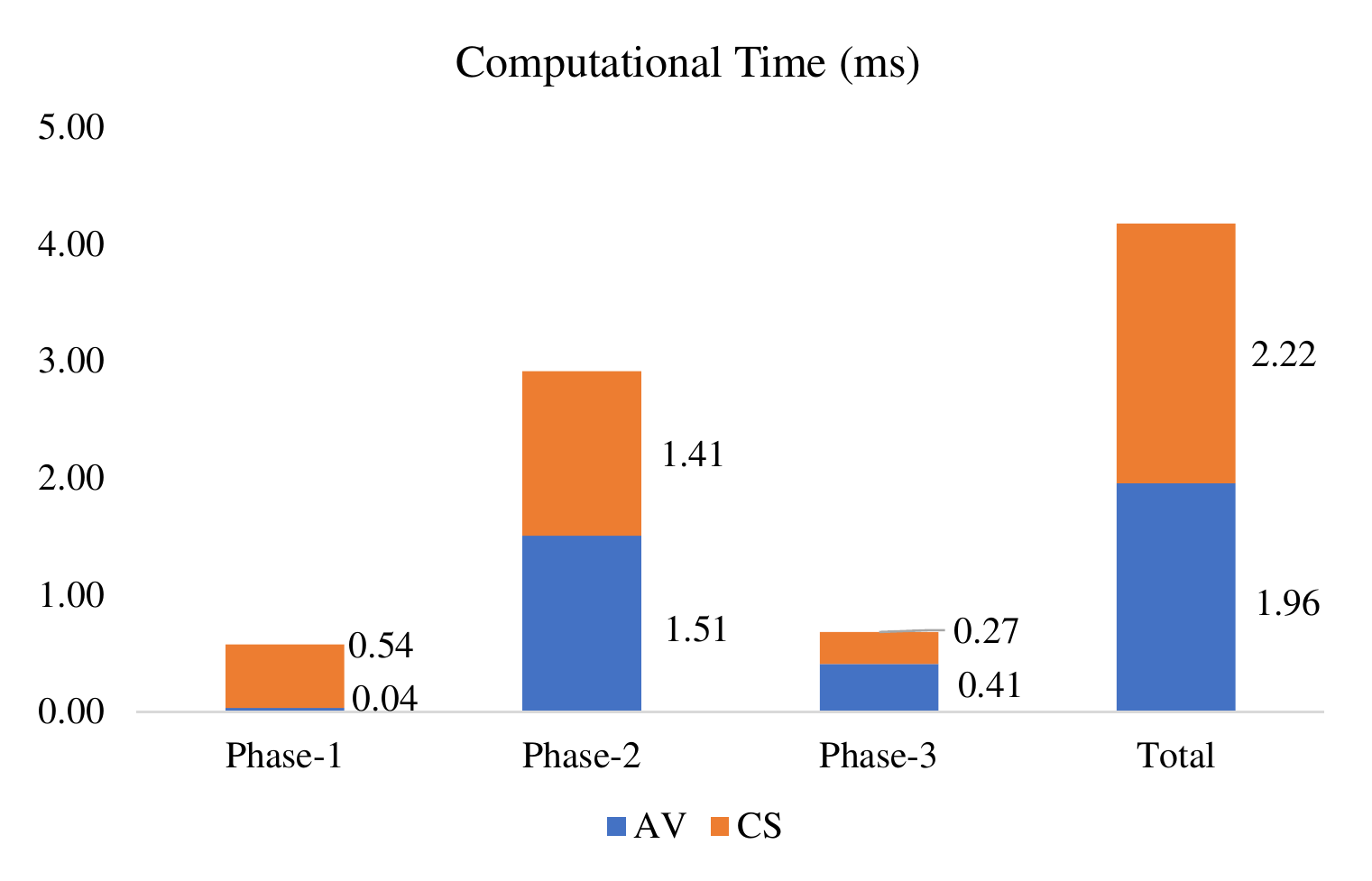}
	\caption{Computational Time of Different Phases of Easy-Sec}
	\label{FIG:computational}
\end{figure}

Now, performance analysis of the proposed Easy-Sec will be shown. As shown in the authentication section, it is divided into three phases. The data presented here are based on 20 collected samples. Total computational time was 4.18ms to complete all the 3 phases. Among 3 phases, phase-2 is required much time due to decryption and other processes. Table \ref{TABLE:Computational} shows computational time of both AV and CS side for all phases. Fig. \ref{FIG:computational} represents computational time requirement for each phases on the proposed Easy-Sec.

\begin{table}[h!]
	\centering
	\caption{Computational Time at both AV and CS}
	\begin{tabular} {l c c c}
		Item & AV Time (ms) & CS Time (ms) & Total Time (ms)\\ [0.5ex]
		\hline
		Phase-1 & 0.04  & 0.54 & 0.58 \\
		Phase-2 & 1.51  & 1.41 & 2.92 \\
		Phase-3 & 0.41  & 0.27 & 0.69 \\
		Total   & 1.96  & 2.22 & 4.18 \\
		\hline
	\end{tabular}
	
	\label{TABLE:Computational}
\end{table}

\begin{figure}[htbp]
	\centering
	\includegraphics[width=0.55\textwidth]{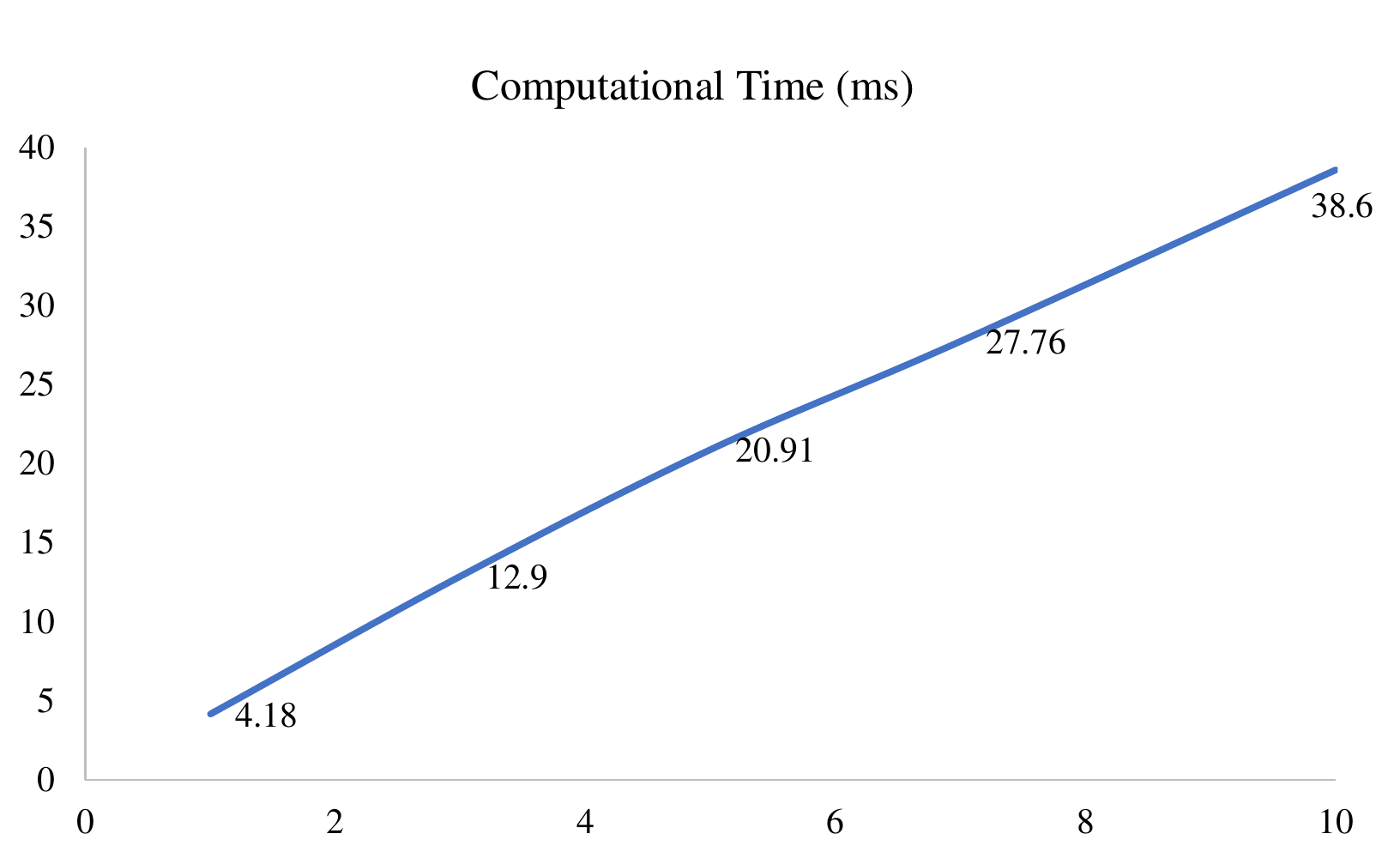}
	\caption{Computational Time of Different Phases of Easy-Sec}
	\label{FIG:scalability}
\end{figure}

Moreover, a network of 10 vehicles (Raspberry Pis) was created to check the computation time and scalability of the proposed framework. Computational time per device does not increase (decrease with a smaller amount) with the increased number of vehicles. The trend of computational time is presented in Fig. \ref{FIG:scalability}.

CRP generation was required for 2 times in the authentication protocol. In this experiment, FPGA was used for responses generation after feeding challenges from Raspberry PI. Table \ref{TABLE:Communication_Time} presents communication time between Raspberry PI \& FPGA and response generation time. As responses will be generated in the same SoC, only response generation time is considered in computational time.

\begin{table}[h!]
	\centering
	\caption{CRP Generation and Communication Time}
	\begin{tabular} {l c}
		Item & Time (ms)  \\ [0.5ex]
		\hline
		Response Generation &  0.4 \\
		Raspberry PI and FPGA Communication &  35.0 \\
		Total &  35.4 \\
		\hline
	\end{tabular}
	
	\label{TABLE:Communication_Time}
\end{table}

\begin{figure}[htbp]
	\centering
	\includegraphics[width=0.75\textwidth]{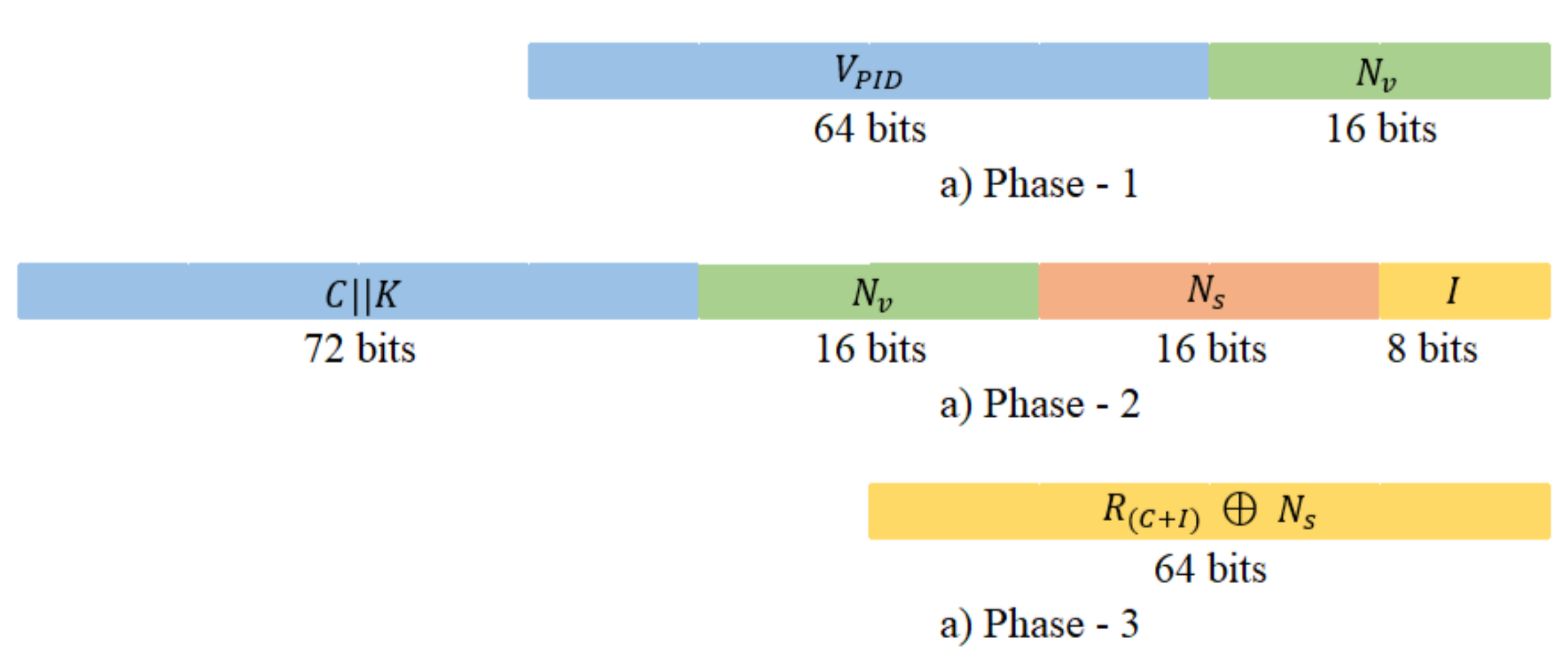}
	\caption{Message flow during each phase of Easy-Sec}
	\label{FIG:Overhead_results}
\end{figure}

For calculating communication overhead in the current work, 64 bits were used as $PID$ and random nonce was considered as 16 bits. Fig. \ref{FIG:Overhead_results} shows the distribution of message flow of each step. 

Table \ref{TABLE:Communication_Overhead} shows communication overhead of the proposed Easy-Sec.  

\begin{table}[h!]
	\centering
	\caption{Communication Overhead}
	\begin{tabular} {l c}
		Item & Communication Overhead (bytes)  \\ [0.5ex]
		\hline
		Phase-1 &  10 \\
		Phase-2 &  14 \\
		Phase-3 &  8 \\
		Total & 32 \\
		\hline
	\end{tabular}
	
	\label{TABLE:Communication_Overhead}
\end{table}

As per data flow, total communication overhead of the proposed scheme is 32 bytes. If SHA-256 is used in the encryption process, then total overhead will be 59 bytes. Communication overhead depends on $PID$ length, types of encryption, random nonce length, size of $K$-bit etc. Hence, final communication overhead will be based on selection of above parameters. Furthermore, for session key update when AV changes RG area, it will take 0.5 ms to extract the new session key as here only decryption of new session key using current session key is required as computation. For communication overhead, it is also low as AV is required to send new $PID$ to new RSU where it will receive encrypted session key. Communication overhead will be as per selection of $PID$ size and encryption mechanism. 

Table \ref{TABLE:Performance} shows the comparative performance analysis among different proposed authentication schemes. From the table, it is evident that the proposed scheme is better than other existing authentication frameworks with respect to performance. Also, the proposed scheme is secured against known security threats which are shown in the security analysis section. 

\begin{table}[h!]
	\centering
	\caption{Performance Comparison}
	\begin{tabular} {p{2cm} p{2cm} p{2cm} p{2cm} p{5cm}}
		Item & Year & Communication Overhead (bytes) & Computational Cost (ms) & Remarks \\ [0.5ex]
		\hline
		Li et al. \cite{li2019cl} & 2019 & 408 & 0.5* & Only message generation and verification are considered; High performance Xeon CPU is used instead of OBU for simulation \\
		Han et al. \cite{han2021implementing} & 2021 & ** & 15.7 & Should have higher overhead due to sending multiple parameters\\
		Thumbur et al. \cite{thumbur2020efficient} & 2020 & 184 & 27 & Many cryptographic operations\\
		Alladi et al. \cite{alladi2020lightweight} & 2020 & ** & 22 & Communication overhead should be higher as it sends keys, IDs, nonce, timestamps etc few times\\
		Aman et al. \cite{aman2020privacy} & 2020 & 24*  & ** & Communication overhead is 102 bytes by calculating as this paper \\
		Javaid et al. \cite{javaid2020scalable} & 2020 & 512 & ** & It is a blockchain-based authentication scheme \\
		This Paper & & 32 & 5 & Simple scheme with low cost and overhead\\
		\hline
		\multicolumn{5}{l}{{*} - Full authentication process is not considered} \\
		\multicolumn{5}{l}{{**} - Information is not provided} \\
		\hline
	\end{tabular}
	
	\label{TABLE:Performance}
\end{table}

\subsection{Formal Analysis of Security Properties}
Formal analysis of the proposed authentication scheme, Easy-Sec, will be depicted in this section. Burrows, Abadi and Needham (BAN) logic \cite{burrows1989logic} \cite{deebak2020smart} will be used to evaluate mutual authentication process. 

\subsubsection{Notations}
To illustrate the significance of each interference postulated in BAN logic, the basic notations and corresponding descriptions which are used as follows.

\begin{itemize}
	\item \textit{P believes X (P $|$$\equiv$ X )}: This formula states that X is true and P believes X, or P would be entitles to believe X. This formula is considered to be the central identifier of the BAN logic..
	\item \textit {P sees X (P $\triangleleft$ X)}: X is the part of the message received by P; P can read and repeat the message or X.
	\item \textit{P once said X (P $|$$\sim$ X)}: Among all the transmitted messages of entity P, a message contains X but it is not possible to identify whether the transmission of the message in the current data flow of the protocol or it was transmitted long time before. However, it is known for the current execution time that P believe X.
	\item \textit{P controls X (P $|$$\Rightarrow$ X)}: This statement indicates that the node P has complete jurisdiction/authorization over the statement X and it should be considered that this complete control access over X provides trustworthy. For example, a server or central database is considered as a trusted entity or trusted third party for generation and management of encryption keys.
	\item \textit{Fresh X (\#(X))}: Usually, nonces are used to make a message X as fresh where it was not processed before the current run of the protocol. 
	\item \textit{P has complete control over X (P $_{\Rightarrow}^{\triangleleft}$X))}: Complete control over the message X by P is considered by this statement. Generally, it is applied when the central body or authority will decide that it is recommended to apply.
	\item \textit{X or Y in a formula ($<$X,Y$>$)}: it is used when the information X or Y is a part of a formula.
	\item \textit{Secret between P and Q (P 	$_{\rightleftharpoons}^{X}$ Q)}: This formula is used to represent that the secret key or formula X is only between the entities P and Q. For instance, this is used as a password/key to prove identity to each other.
	\item \textit{Shared key between P and Q (P 	$_{\leftrightarrow}^{K}$ Q)}: This formula indicates that shared K is only discovered by the entity P and Q. Not other entity can discover the shared key K.
	\item \textit{Encryption of message X using key K ($\{$X$\}_K$)}: The message of formula X is encrypted using the key K is presented by the notation. The formula is used to protect the message or formula from unauthorized entity. 
\end{itemize}

\subsubsection{Inference Rules} Different sets of inference rules are used in the BAN logic to illustrate the significance/justification of the messages. The concluding remarks of inference rules are as follows:
\begin{itemize}
	\item [$\circ$] \textit {${IR}_{1}$}: $<$Nonce-Verification Rule$>$ If P believes that $X$ is random and new and that it was said by entity Q, P considers that Q believes $X$ as well.
	\begin{equation*} \label{eq1}
		\begin{split}
			\frac{P |\equiv \#(X), P|\equiv S |\equiv | \sim X} {P|\equiv S |\equiv X}
		\end{split}
	\end{equation*}
	\item [$\circ$] \textit {${IR}_{2}$}: $<$Jurisdiction Rule$>$ If P trusts that Q has absolute jurisdiction over $X$, and Q believes $X$ as well, P believes Q on the truth of $X$.
	\begin{equation*} \label{eq1}
		\begin{split}
			\frac{P |\equiv P \Rightarrow X, P |\equiv S |\equiv X} {P|\equiv X}
		\end{split}
	\end{equation*}
	\item [$\circ$] \textit {${IR}_{3}$}: $<$Key Freshness Rule$>$ If one portion of the formula is fresh, the complete formula will be fresh. If $X$ is fresh, then the entity P will believe in the freshness of $(X,Y)$
	\begin{equation*} \label{eq1}
		\begin{split}
			\frac{P |\equiv \#(X)} {P|\equiv \#(X,Y)}
		\end{split}
	\end{equation*}
	\item [$\circ$] \textit {${IR}_{4}$}: $<$Belief Rule$>$ The entity P will believe a statement if it believes all the components of the statement separately. If the statement $(X, Y)$ is trusted by the entity P, then it will believe statement $X$.
	\begin{equation*} \label{eq1}
		\begin{split}
			\frac{P |\equiv (X,Y)} {P|\equiv (X)}
		\end{split}
	\end{equation*}
	\item [$\circ$] \textit {${IR}_{5}$}: $<$Secret Sharing Rule$>$ In either way, the same key is utilized between a couple of entities.
	\begin{equation*} \label{eq1}
		\begin{split}
			\frac{P |\equiv Q |\equiv R \overset{K}{\rightleftharpoons} R^{'}} {P |\equiv Q |\equiv R^{'} \overset{K}{\rightleftharpoons} R}
		\end{split}
	\end{equation*}
	\item [$\circ$] \textit {${IR}_{6}$}: $<$Shared Key Rule$>$
	If a message $X$ is encrypted using key $K$ and the entity P trusts that the key $K$ is shared with the entity Q, then P will believe the message $X$ was once stated by entity Q.
	\begin{equation*} \label{eq1}
	\begin{split}
		\frac{P |\equiv Q \overset{K} {\leftrightarrow} P, P   {\triangleleft} \{X\}_{K} } {P |\equiv Q |{\sim} X}
	\end{split}
	\end{equation*}
\end{itemize}

\subsubsection{Initial Assumptions} To mutual authenticate the security properties, following assumptions are considered. Also, RSU and RG aren't considered in the security authentication process as these are mutually authenticated earlier. 
\begin{itemize}
	\item [$\circ$] \textit {${A}_{1}$}: The CS stores the response $R$ for challenge $C$. So, the	CS has the $K$-bit response and the CS believes that it is between the CS and the AV. \\	{CS $|$$\equiv$  CS $_{\rightleftharpoons}^{R^K}$ AV} \\
	\item [$\circ$] \textit {${A}_{2}$}: The AV can generate response $R$ and $R^K$ for challenge $C$. The AV trusts that the secret response is between the CS and the AV.	\\	{AV $|$$\equiv$  CS $_{\rightleftharpoons}^{R^K}$ AV} \\
	\item [$\circ$] \textit {${A}_{3}$}: The CS has the response for the challenge $(C+I)$ and the CS believes that it is secret between the CS and the AV.\\	{CS $|$$\equiv$  CS $_{\rightleftharpoons}^{R_{C+I}}$ AV} \\
	\item [$\circ$] \textit {${A}_{4}$}: The AV assumes that the response for challenge $(C+I)$ is kept between the CS and the AV. Also, the AV is able to generate response for challenge $(C+I)$.	\\	{AV $|$$\equiv$  CS $_{\rightleftharpoons}^{R_{C+I}}$ AV} \\
\end{itemize}

\subsubsection{Idealized Form} Followings are showed the idealized form of the messages of the proposed Easy-Sec framework.

\begin{itemize}
	\item [$\circ$] \textit {${I}_{1}$}: AV $\to$ CS: The AV sends the statement of $PID$ and $N_v$. The statement is fresh as $N_v$ is fresh.\\ \{$V_{PID}$, $N_{V}$, \#($N_{V}$)\} \\
	\item [$\circ$] \textit {${I}_{2}$}: CS $\to$ AV: The statement from the CS to the AV is fresh as the statement contains fresh $C$, $K$, and $I$.\\ \{$C, K,$  \{$N_{v}$, $N_{s}$, I$\}_{R^{K+1}}$, \#$(C, K, I)$\} \\
	\item [$\circ$] \textit {${I}_{3}$}: AV $\to$ CS: Fresh response of challenge $(C+I)$ and fresh $N_s$ make sure the shared statement is fresh.\\ \{$R_{C+I}$, $N_{s}$, \#($R_{C+I}$, $N_{s}$) \} \\
\end{itemize}

\subsubsection{Goals of Proposed Easy-Sec Framework} It is required to achieve below two goals to complete the mutual authentication of the proposed protocol.
\begin{itemize}
	\item [$\circ$] \textit {${G}_{1}$}: The first goal is to ensure that the secret key ($(K+1)$ bit response of challenge $C$) is only discovered by the entity CS and AV.	\\	{CS $|$$\equiv$ AV $|$$\equiv$ $<$CS $_{\longleftrightarrow}^{R^{K+1}}$ AV$>$} \\
	\item [$\circ$] \textit {${G}_{2}$}: The second goal is to ensure the response of challenge $(C+I)$ between the CS and the AV.\\{CS $|$$\equiv$ AV $|$$\equiv$ $<$CS $_{\longleftrightarrow}^{R_{C+I}}$ AV$>$} \\
\end{itemize}

\subsubsection{Formal Verification Proof} Now, mutual authentication of the work Easy-Sec will be verified based on the above mentioned inference rules, initial assumptions, idealized form and goals. The detailed processes are followings: 

\begin{itemize}
	\item [$\circ$] \textit {${FV}_{1}$}: From ${I}_{1}$ and by practicing ${IR}_{1}$ and ${IR}_{3}$, equation (1) is desired to be obtained which is shown below as below:	
	\begin{equation} \label{eq2}
		\begin{split}
			\frac{CS |\equiv \# <N_v>, CS |\equiv MD |\equiv \#(N_v)} {AV |\equiv \#(N_v)}
		\end{split}
	\end{equation}
	\item [$\circ$] \textit {${FV}_{2}$}: From ${I}_{2}$ and by practicing ${IR}_{1}$, ${IR}_{3}$ and ${IR}_{6}$, it is desired to obtain (2) and achieves goal ${G}_{1}$:
	\begin{equation*} 
	{AV |\equiv\# (C,K), AV |\equiv CS \overset{R^{K+1}} {\leftrightarrow} AV, AV {\triangleleft} \{N_v,N_s,I\}_{R^{K+1}},} 
	\end{equation*}	
	\begin{equation} \label{eq1}
		\begin{split}
			\frac{AV |\equiv CS|\equiv \sim (C)} {AV |\equiv\# (C,K,N_v,N_s,I), AV|\equiv{R^{K+1}}}
		\end{split}
	\end{equation}
	\item [$\circ$] \textit {${FV}_{3}$}: From ${I}_{3}$ and by practicing ${IR}_{2}$, ${IR}_{4}$ and ${IR}_{5}$, it is desired to obtain (3) which achieves the goal ${G}_{2}$.
	\begin{equation} \label{eq1}
	\begin{split}
		\frac{CS |\equiv AV | _{\Rightarrow}^{\triangleleft} \#(R_{C+I}), CS \overset{\#(R_{C+I})}{\rightleftharpoons} AV, CS|\equiv AV|\equiv \sim (N_s) } {CS |\equiv R_{C+I}}
	\end{split}
	\end{equation}
\end{itemize}

\subsection{Informal Analysis of Security Properties}
This section will show how the proposed authentication scheme Easy-Sec is able to take security measures considering an adversary has the ability to modify and overhear the transmitted data over public networks.
\subsubsection{Impersonation Attack}
The proposed Easy-Sec considers both client and server impersonation attack. The following shows the process to resist impersonation attack for both cases.
\begin{itemize}
	\item \textit{Server Impersonation Attack:}
	An adversary can impersonate as server if the adversary acts as server to get information from user or if the user tries to communicate with server but it knocks the wrong door. In Easy-Sec, for both cases server needs to send encrypt nonce using $K+1$ bit of a challenge. As any response is not being send either from user or from server, it is not possible for attacker to get response of any challenge. After getting request from the attacker, user will try to decrypt the data using $K+1$ bit response of the challenge. As the adversary encrypted data using other wrong key, user will not be able to decrypt the data and mark as fake server. In this way, user can resist server impersonation attack. Server impersonation attack was performed and result is shown in the Fig. \ref{FIG:Server_Impersonation}.

	\begin{figure}[htbp]
		\centering
		\includegraphics[width=0.65\textwidth ]{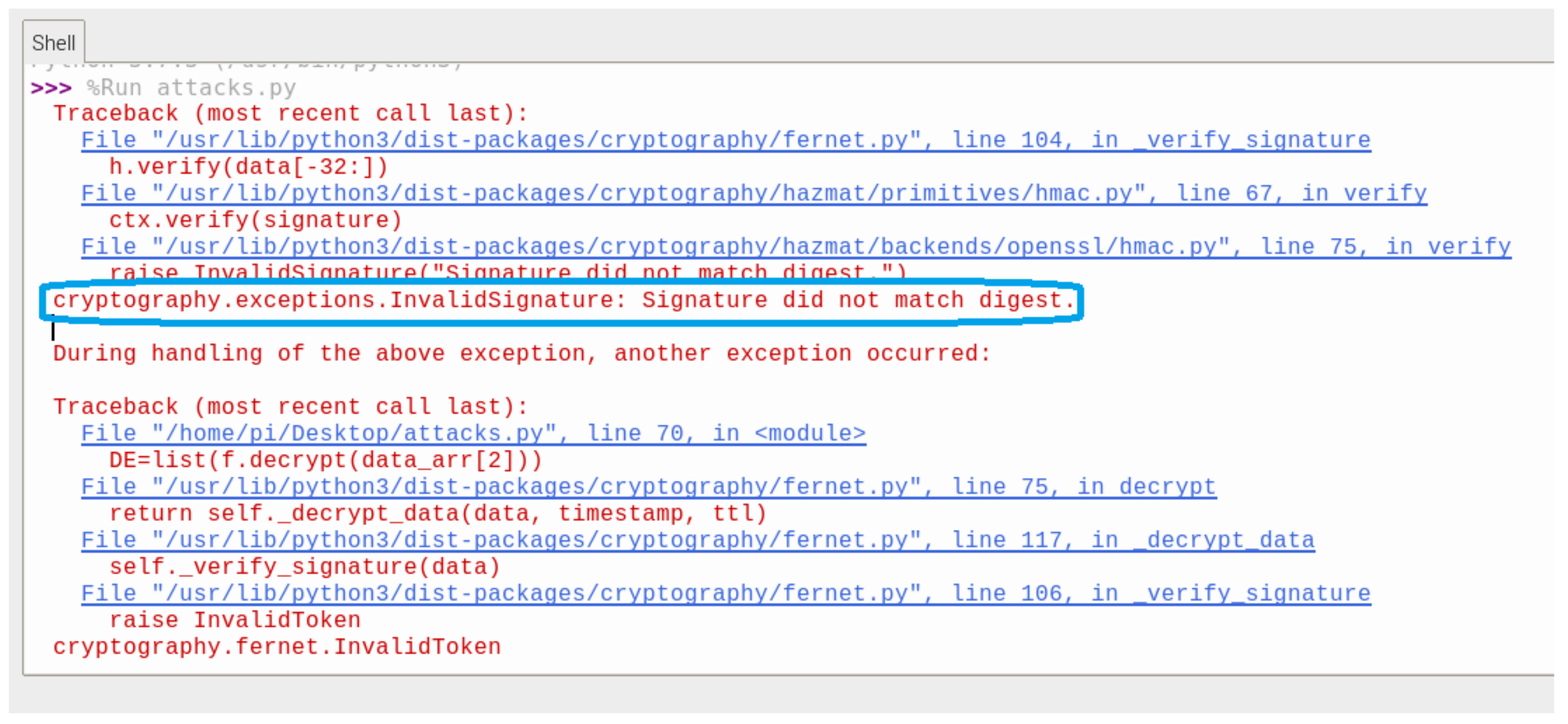}
		\caption{Server Impersonation Attack}
		\label{FIG:Server_Impersonation}
	\end{figure}

	\item \textit{Client Impersonation Attack:}
	Client device does not store any key or password which will be used for authentication, instead of that PUF is used in this work which can generate response on the fly.
	CRP pairs are stored in the secure database of the cloud for authentication purpose. Client device will generate a response based on $C+I$ challenge. As both $I$ and server nonce are encrypted using a variable $K+1$ bit, so adversary can not reveal the random $I$ and also, on the fly regenerated response will be used to perform XOR operation with the nonce of server. Also, To test this, an attacker \textit{A} tried to impersonate a client by sending an XOR result of response and nonce. As both response and nonce were unknown to the adversary, client impersonation attack was failed which is shown in the Fig. \ref{FIG:Client_Impersonation}.
		
	\begin{figure}[htbp]
		\centering
		\includegraphics[width=0.65\textwidth ]{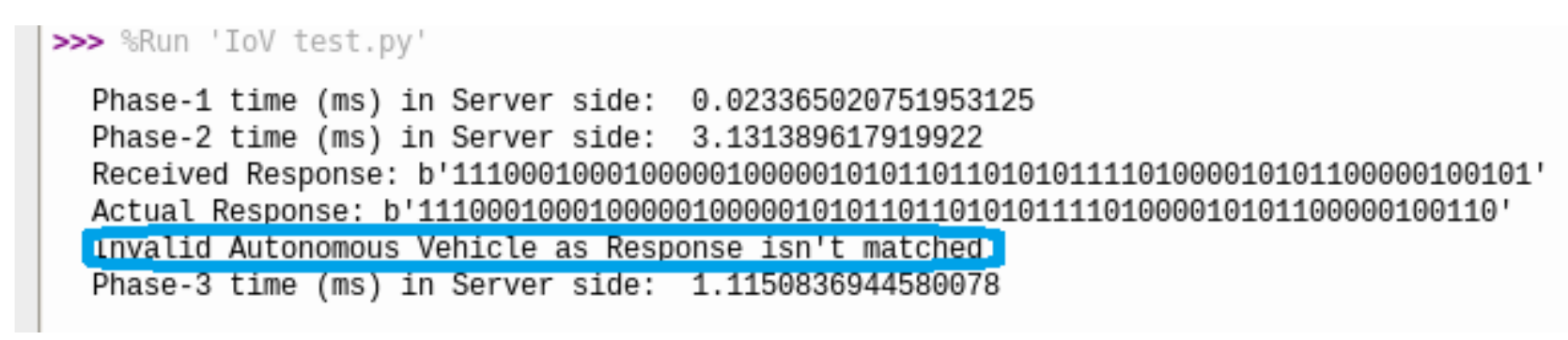}
		\caption{Client Impersonation Attack}
		\label{FIG:Client_Impersonation}
	\end{figure}
	
\end{itemize}

\subsubsection{Side Channel Attack}
Most authentication schemes used stored password, keys in some memory location for authentication purpose which makes the scheme susceptible to side channel attack \cite{siddiqi2020imdfence}. An attacker can use power analysis, timing information etc to extract key bits. To avoid this dependency, PUF is incorporated in this work which which eliminates the requirements of key storage in memory location.
Based on process variation of chips, it will generate R which makes the proposed Easy-Sec resistance against side-channel attack.

\subsubsection{Machine Learning Attack}
\begin{figure}[!h]
		\centering
		\includegraphics[width=.4\textwidth ]{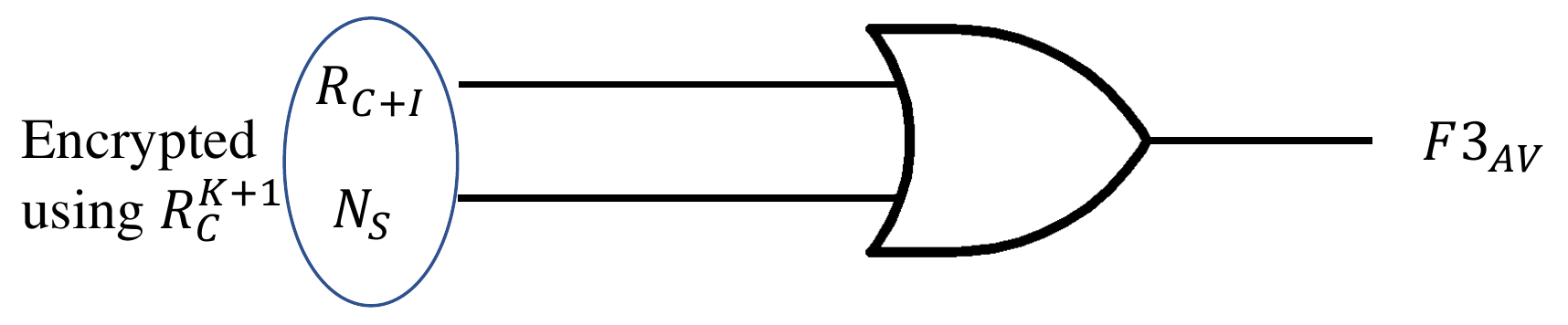}
		\caption{Hiding Response of Client to Resist ML Attack}
		\label{FIG:ML_Attack}
	\end{figure}

By following the pattern of shared key/password/response from a transmitting device, adversary builds a model by introducing machine learning attack to extract shared key/password/response of the entity. In the proposed scheme, a random response using $C+I$ challenge and random nonce $N_{s}$ will be shared using XOR function. As both $I$ and $N_{s}$ are shared using a random encryption, it is not possible for any adversary to guess and find out these. Moreover, response of $C+I$ will be generated on the fly. So, both these input of XOR function is unknown and without knowing any input, it is not possible to extract another input of the XOR function. By following this process, the proposed Easy-Sec can resist machine learning attack. Fig. \ref{FIG:ML_Attack}. shows the XOR operation for sharing response for verification. 
	
\subsubsection{Physical Attack}
Any faulty or physically tampered device can be identified using PUF \cite{gope2020secure}. If an adversary is trying to tamper the controller device of AV or it becomes faulty, then the characteristics of PUF will be changed and it will not be able to accurate response for $C+I$ challenge or $K+1$ bit response for $C$ challenge. As a result, AV will not be able to decrypt message from CS and furthermore, it will fail to send message of phase-3 due to response generation failure. PUF will eliminate the dependency of storing response in some memory location of the AV as it will be generated on the fly. Therefore, an adversary will not be able to gain CRP set even he/she has physical access to the AV. So, it can be said that the proposed Easy-Sec can ensure security against physical attacks as any such tampering attempt would change the behavior of the PUF and the server will be able to detect any CRP anomaly. 

\subsubsection{Dos Attack}
By draining the energy of battery, Dos attack can cause sleep deprivation to make it out of service. The target of the attacker is to shut down the device by forcing to run energy-consuming operations continuously which results in battery depletion. For example, an attacker sends continuous authentication request to the client/server by pretending a valid server/client and client/server will generate response to validate the message. If any device/server gets invalid requests for more than a certain time, for example 5 time, it will place the illegal device in grey list and will not accept further authentication request. So, the proposed scheme is able to resist DoS attack and DoS attack was performed to check the performance.

\subsubsection{Replay Attack}
By sending maliciously or fraudulently repeated or delayed valid data transmission, an attacker can perform replay attack which is also known as playback attack. To prevent reply attack, the clock synchronization and random number mechanism are two mechanisms particularly. However as per \cite{li2019secure}, the clock synchronization is still a research problem in WSN for communication between client node and destination node. Random number mechanism, random nonce of AV $N_{V}$ and CS $N_{S}$, is adopted in this scheme to resist the replay attack. Furthermore, instead of storing new PID, $PID_{New}$, directly it is being sent by adding 1 by AVs to protect against replay of the previous message \cite{siddiqi2020imdfence} and $PID_{New}$ is only valid for certain time. Considering all these, it can be stated that the Easy-Sec can block replay attack by using random number and this work in not affected by clock synchronization problem.

\subsubsection{Eavesdropping Attack}
Characteristics of PUF shows that each device is unique properties which can used for secure communication between AV and CS by unique CRP set. Since all the CRPs are stored in the CS during registration phase, AV generates on the fly and these are used to encrypt message which makes the eavesdroppers to resist to impersonate the messages  \cite{yildiz2020plgakd}. 

\subsubsection{Man In the Middle Attack}
Different secret keys, which is responses, and different bit length are used in message transmission for the authentication part in phase-2 and 3. Therefore, the AV and CS both will generate different messages, which prevents MITM for the authentication part. All responses which are $R^{K+1}$ and $R_{C+I}$ are not stored anywhere in the AV and are generated on the fly on nodes’ PUF for MITM attack resistance. Hence, inside intruders will be get any chance to perform the MITM attack without knowing CRPs, XOR functions and random numbers ($K$-bit and nonces) \cite{yildiz2020plgakd}.

\subsubsection{Anonymous Identity}
In this work, both privacy and data integrity are focused to protect those. To maintain privacy preservation by hiding sensitive information or files of the AV to the CS through RSU and RG. Pseudo-random identifiers are used instead of real id to avoid exposing real identifiers so that any attacker cannot identify a user according to the associated AV. Furthermore, in each authentication process, a new pseudo-random identifier {$PID_{New}$ will be assigned to the AV in the encrypted form. 

\subsubsection{Forward Secrecy}
The main purpose of forward secrecy is to ensure that previously established session keys remain secure in the case of the keys are leaked. In the proposed Easy-Sec, the session key/response is depends on the random challenge $C$, random $K$-bit and a random increment of challenge $I$. As these numbers are not fixed or follow a pattern, it can be concluded that the proposed protocol guarantees forward secrecy.
	
The above discussions depicts that the robustness of the proposed authentication framework - Easy-Sec and how it is able to resist known security threats. Now, in the Table II, comparison among different proposed protocols in the literature review are shown with respect to resistance against security threats.

\section{Conclusion and Future Directions}
\label{SEC:Conclusion}
In this paper, a secure and ultralight authentication scheme is proposed for verifying the AV to send/receive traffic information and firmware update process. It is an efficient and secure PUF-based authentication framework for the application of the IoV. The proposed scheme is free from complex certificate management, key storage problem. As the authentication of both CS and AV is done by using two message flow, this technique simplifies the verification time, computational cost, communication overhead, bandwidth requirement and storage space at AV, RSU and RG. The proposed authentication protocol is shown to prevent from disrupting the security features by both informal security analysis and BAN logic. Performance analysis shows that the proposed authentication scheme is more efficient compared to similar state-of-the authentication schemes in terms of security, computational, and communication point of view. Hence, the proposed PUF-based authentication scheme is more feasible for the IoV environment. 

Though this paper shows very low computational time and communication overhead, it will be targeted to further lowering time and cost by reducing CRP generation timeline and other optimization schemes. Moreover, blockchain will be introduced to decentralize and compare the performance.

\bibliographystyle{unsrt}
\bibliography{Bibliography_PUF-Authentication-for-IoV}

 \begin{wrapfigure}[8]{l}{25mm} 
	\includegraphics[height=1.1in]{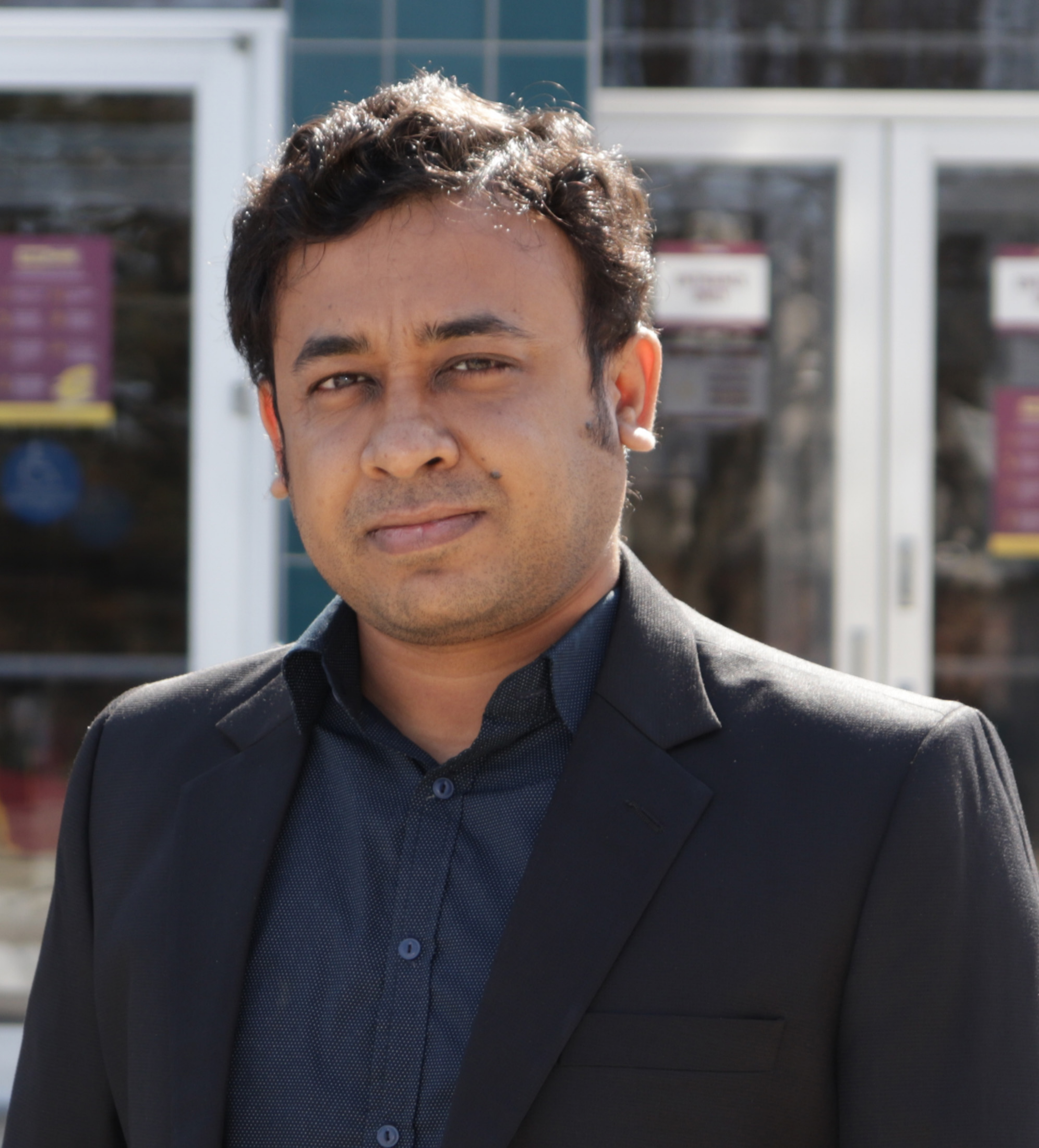}
\end{wrapfigure}\par
\textbf{Pintu Kumar Sadhu} received received the
Bachelor of Science degree in Electrical \& Electronic Engineering from the Khulna University Of Engineering \& Technology, Bangladesh in 2010. He worked as a specialist in Grameenphone Ltd., Bangladesh from 2012 to 2020. He is currently doing Master of Science in Hardware Assisted Security Systems Laboratory, at Central Michigan University. He is the author of two articles published in peer reviewed conference proceedings. His research interests are providing Hardware Assisted Security for Internet of Things, Internet of Vehicles, Machine Learning, and Embedded Systems.
\\
\begin{wrapfigure}[10]{l}{25mm} 
	\vspace{-\baselineskip}
\includegraphics[height=1.5in]{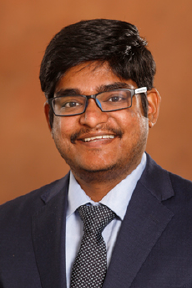}
\end{wrapfigure}\par

\textbf{Venkata P. Yanambaka} (M’19) received the Bachelor of Technology degree in electronics and communications from the Jawaharlal Nehru Technological University, India, in 2014 and the Ph.D. degree from the Smart Electronic Systems Laboratory, Department of Computer Science and Engineering, University of North Texas. He is currently an Assistant Professor with the School of Engineering and Technology, Central Michigan University. He has authored 30 research articles which include multiple journals/transactions articles. His research interests are in security in Internet of Things, energy-efficient circuits and systems, and application-specific systems design. He is the director of Hardware Assisted Security Systems Laboratory at Central Michigan University. He is the Treasurer for IEEE Northeast Michigan Section.
\\

\begin{wrapfigure}[10]{l}{25mm} 
	\vspace{-\baselineskip}
\includegraphics[height=1.5in]{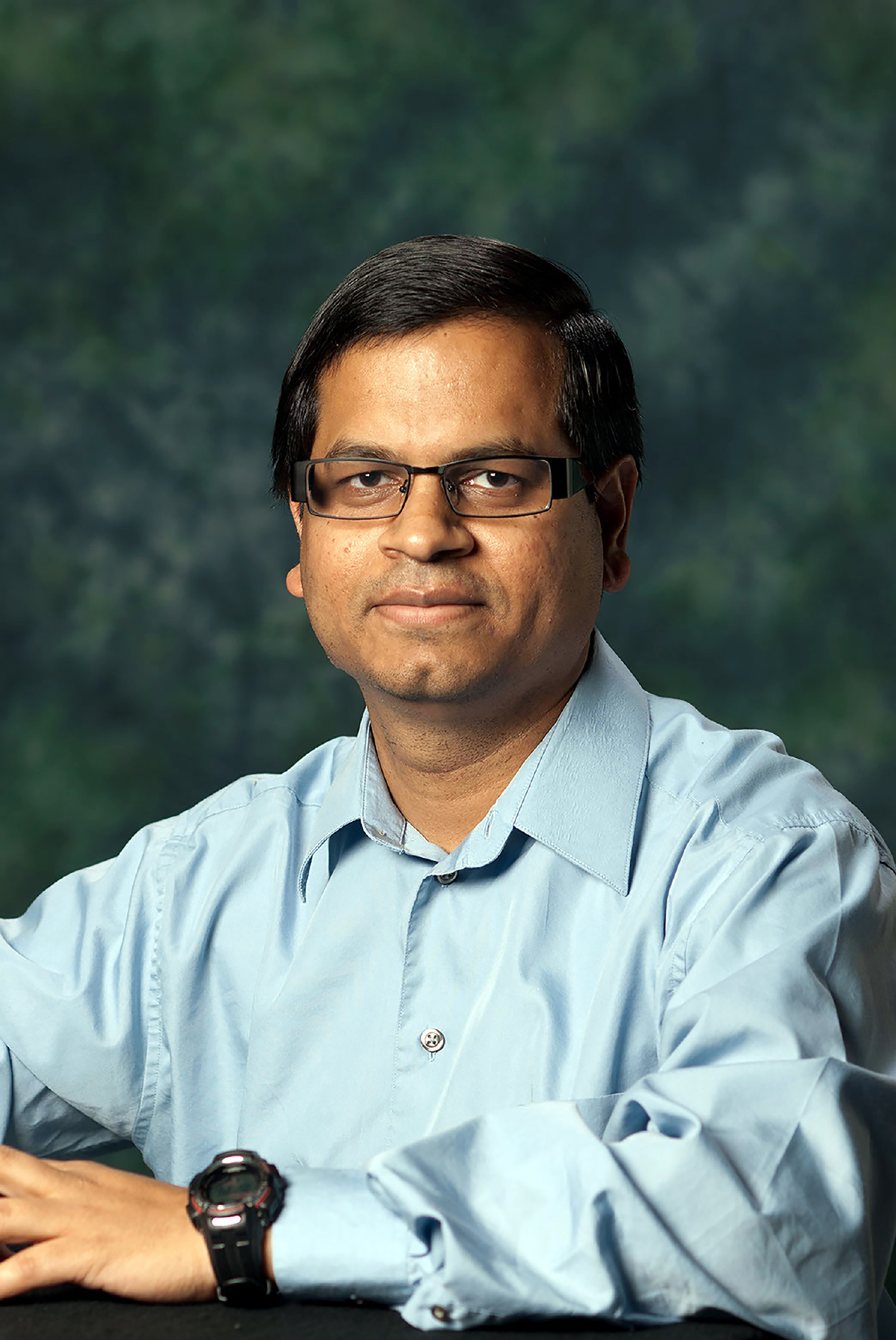}
\end{wrapfigure}\par
\textbf{Saraju P. Mohanty} received the bachelor’s degree (Honors) in electrical engineering from the Orissa University of Agriculture and Technology, Bhubaneswar, in 1995, the master’s degree in Systems Science and Automation from the Indian Institute of Science, Bengaluru, in 1999, and the Ph.D. degree in Computer Science and Engineering from the University of South Florida, Tampa, in 2003. He is a Professor with the University of North Texas. His research is in ``Smart Electronic Systems'' which has been funded by National Science Foundations (NSF), Semiconductor Research Corporation (SRC), U.S. Air Force, IUSSTF, and Mission Innovation. He has authored 400 research articles, 4 books, and 7 granted and pending patents. His Google Scholar h-index is 45 and i10-index is 180 with 8600 citations. He is regarded as a visionary researcher on Smart Cities technology in which his research deals with security and energy aware, and AI/ML-integrated smart components. He introduced the Secure Digital Camera (SDC) in 2004 with built-in security features designed using Hardware-Assisted Security (HAS) or Security by Design (SbD) principle. He is widely credited as the designer for the first digital watermarking chip in 2004 and first the low-power digital watermarking chip in 2006. He is a recipient of 14 best paper awards, Fulbright Specialist Award in 2020, IEEE Consumer Technology Society Outstanding Service Award in 2020, the IEEE-CS-TCVLSI Distinguished Leadership Award in 2018, and the PROSE Award for Best Textbook in Physical Sciences and Mathematics category in 2016. He has delivered 15 keynotes and served on 13 panels at various International Conferences. He has been serving on the editorial board of several peer-reviewed international transactions/journals, including IEEE Transactions on Big Data (TBD), IEEE Transactions on Computer-Aided Design of Integrated Circuits and Systems (TCAD), IEEE Transactions on Consumer Electronics (TCE), ACM Journal on Emerging Technologies in Computing Systems (JETC), and Springer Nature Cpmputer Science (SN-CS). He was the Editor-in-Chief of the IEEE Consumer Electronics Magazine (MCE) during 2016-2021. He served as the Chair of Technical Committee on Very Large Scale Integration (TCVLSI), IEEE Computer Society (IEEE-CS) during 2014-2018 and on the Board of Governors of the IEEE Consumer Electronics Society during 2019-2021. He serves on the steering, organizing, and program committees of several international conferences. He is the steering committee chair/vice-chair for the IEEE International Symposium on Smart Electronic Systems (IEEE-iSES), the IEEE-CS Symposium on VLSI (ISVLSI), and the OITS International Conference on Information Technology (OCIT). He has mentored 2 post-doctoral researchers, and supervised 13 Ph.D. dissertations, 26 M.S. theses, and 11 undergraduate projects.

\begin{wrapfigure}[10]{l}{25mm} 
	\vspace{-\baselineskip}
	\includegraphics[height=1.5in]{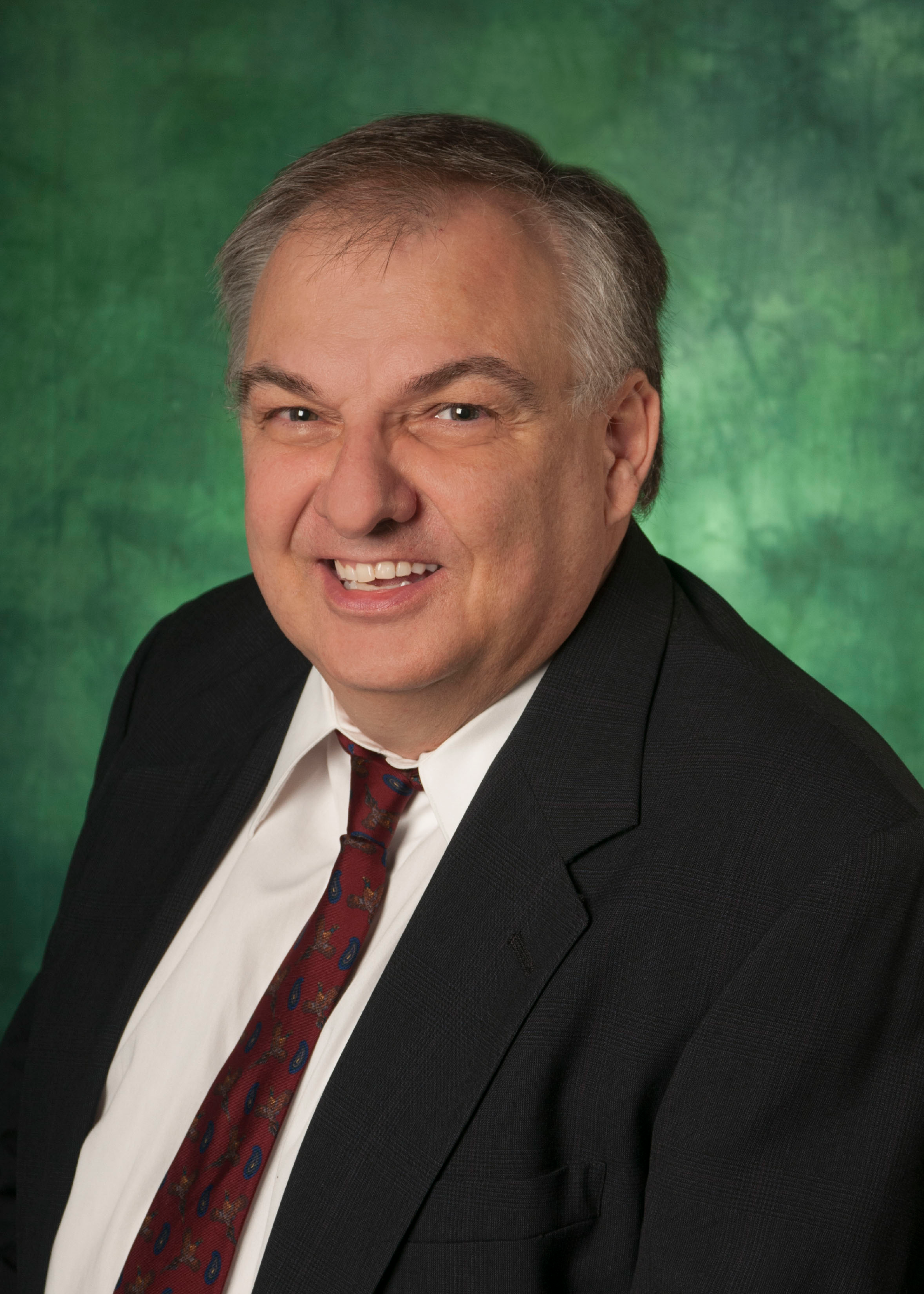}
\end{wrapfigure}\par
\textbf{Elias Kougianos} (SM’07) received a BSEE from the University of Patras, Greece in 1985 and an MSEE in 1987, an MS in Physics in 1988 and a Ph.D. in EE in 1997, all from Louisiana State University. From 1988 through 1997 he was with Texas Instruments, Inc., in Houston and Dallas, TX. Initially he concentrated on process integration of flash memories and later as a researcher in the areas of Technology CAD and VLSI CAD development. In 1997 he joined Avant! Corp. (now Synopsys) in Phoenix, AZ as a Senior Applications engineer and in 2001 he joined Cadence Design Systems, Inc., in Dallas, TX as a Senior Architect in Analog/Mixed-Signal Custom IC design. He has been at UNT since 2004. He is a Professor in the Department of Electrical Engineering, at the University of North Texas (UNT), Denton, TX. His research interests are in the area of
Analog/Mixed-Signal/RF IC design and simulation and in the development of VLSI architectures for multimedia applications. He is an author of over 120 peer-reviewed journal and conference publications.

\end{document}